\documentclass[aps,dvips]{revtex4}
\usepackage{amsmath}
\usepackage{amsfonts}
\usepackage{amssymb}
\usepackage{graphicx}

\begin{document}

\title{Electron-positron annihilation into $\phi f_{0}(980)$ and clues for a
new $1^{--}$ resonance}
\author{M. Napsuciale$^{1,2}$, E. Oset$^{1}$, K. Sasaki$^{1}$, C. A.
Vaquera-Araujo$^{1,2}$}
\affiliation{$^{1}$ Departamento de F\'{\i}sica Te\'{o}rica and Instituto de F\'{\i}sica
Corpuscular, Centro Mixto Universidad de Valencia-CSIC, 46000 Burjassot,
Valencia, Spain}
\affiliation{$^{2}$ Instituto de F\'{\i}sica, Universidad de Guanajuato, Lomas del Bosque
103, Fraccionamiento Lomas del Campestre, 37150, Le\'{o}n, Guanajuato, M\'{e}%
xico}

\begin{abstract}
We study the $e^{+}e^{-}\rightarrow\phi\ \pi\ \pi$ reaction for pions in an
isoscalar s-wave which is dominated by loop mechanisms. For kaon loops we
start from the conventional $R\chi PT$, but use the unitarized amplitude for 
$K\overline{K}-\pi\pi$ scattering and the full kaon form factor instead of
the lowest order terms. We study also effects of vector mesons using $R\chi
PT$ supplemented with the conventional anomalous term for $VVP$ interactions
and taking into account the effects of heavy vector mesons in the $K^{\ast}K$
transition form factor. We find a peak in $m_{\pi\pi}$ around the $%
f_{0}(980) $ as in the experiment. Selecting the $\phi f_{0}(980)$
contribution as a function of the $e^{+}e^{-}$ energy we also reproduce the
experimental data except for a narrow peak, yielding support to the
existence of a $1^{--}$ resonance above the $\phi f_{0}(980)$ threshold,
coupling strongly to this state.
\end{abstract}

\maketitle

\section{Introduction}

The initial state radiation $e^{+}e^{-}\rightarrow\gamma_{ISR}+\gamma^{\ast
}\rightarrow\gamma_{ISR}+X$ in electron-positron machines is being used to
study electron-positron annihilation into hadronic states $X$, scanning
energies below the original design in the so-called radiative return method.
This method has proved to be useful both in the study of the properties of
low lying resonances in $\phi$ factories \cite{daphne} as well as in the
measurement of the cross section for electron-positron annihilation into
different hadronic final states in $B$ factories \cite{BaBar}. In the latter
case it is possible to study electron-positron annihilation into hadronic
states over the range from $1\ GeV$ up to $5\ GeV$ with a clean
identification of the desired final states over the hadronic background.
Detailed analysis of some of these processes show enhancements of the
corresponding cross sections whose proper description seems to require the
existence of new resonances. Indeed, a broad structure was found in the $%
e^{+}e^{-}\rightarrow\gamma _{ISR}J/\psi\pi^{+}\pi^{-}$ cross section
showing the existence of a resonance with a mass of about $4.26\ GeV$ \cite%
{BBY}. More recently, in studying the cross section as a function of the
center of mass for $e^{+}e^{-}\rightarrow\gamma_{ISR}\phi\pi\pi$ with the
dipion mass close to the $f_{0}(980),$ another structure was found around $%
2.2\ GeV$ indicating the existence of a new resonance with a mass of about $%
2.175\ GeV$ and a width of $58\ MeV$ \cite{BBX}.

For final pions in a $C$ even state, the leading electromagnetic
contributions to the $e^{+}e^{-}\rightarrow\phi\pi\pi$ process come from the
exchange of a virtual photon. The quark lines of the $\phi$ and $\pi\pi$
final states are disconnected thus at tree level the $\gamma^{\ast}%
\rightarrow\phi\pi\pi$ can only be induced by sequential decays like $%
\gamma^{\ast}\rightarrow\omega \pi\pi\rightarrow\phi\pi\pi$ which are
suppressed by the small $\omega-\phi$ mixing. We explored this possibility
finding this contribution rather small. The natural mechanisms appear at one
loop level. In particular for a dipion mass close to the $f_{0}(980)$ this
process involves the $\gamma^{\ast}\phi f_{0}$ vertex function with a photon
with a virtuality above $2\ GeV$. The very same vertex function appears also
in one of the mechanisms (dominant in the case of neutral pions) for the
radiative decay $\phi\rightarrow\pi \pi\gamma$ recently measured in
electron-positron $\phi$ factories \cite{KLOE} but there photons are
on-shell. The vertex function at $k^{2}=0$ , the $\phi f_{0}\gamma$
coupling, appearing in these decays is an important piece in the elucidation
of the structure of the lowest lying scalar nonet.

The $\phi\rightarrow\pi\pi\gamma$ decays have been studied in effective
models for non-perturbative QCD \cite{MSL} incorporating scalar degrees of
freedom and in unitarized chiral perturbation theory \cite{MHOT} (see also
applications to $\phi\rightarrow K^{0}\overline{K^{0}}\gamma$ in \cite{Oller}%
). In both formalisms, the dynamics is dominated by the chain $%
\phi\rightarrow S\gamma\rightarrow\pi\pi\gamma$ where the $\phi\rightarrow
S\gamma$ decay is induced at one loop level through charged kaon loops which
couple to the explicit scalar fields in the former case or generate them
dynamically through $K\overline{K}-\pi\pi$ rescattering in the latter case.
The very same dynamics must be at work in the case of virtual photons and
should be the dominant one for low photon virtualities. The calculation of
such effects is the subject of this paper.

Unlike the case of the $\phi\rightarrow S$ $\gamma$ decay where the real
photon tests only the electric charge, here we have a highly virtual photon
which couples to higher multipoles and the way to incorporate systematically
the effects of kaon loops is to consider the full kaon form factor $%
F_{K^{+}}(k^{2})$ in the $\gamma K^{+}K^{-}$ interaction. Furthermore,
although the contribution of neutral kaons vanishes for real photons, in the
case of virtual photons the $\gamma^{\ast}K^{0}\overline{K^{0}}$ coupling is
not null and we must consider also neutral kaon loops with the corresponding
form factor. The challenge here is the proper characterization of the kaon
form factor at the energy of the reaction. Fortunately we have at our
disposal both a theoretical calculation of the neutral and charged kaon form
factors in $U\chi PT$ \cite{OOP} and direct measurements \cite{BiselloFF} in
the energy region of interest. In the former case, the kaon form factor is
matched with the perturbative QCD predictions at high energy and to $\chi PT$
at low energy and, although the calculated form factor cannot account for
the effects of excited vector mesons lying around $1.6$ $GeV$, it is in
agreement with the scarce experimental data above $2$ $GeV$. Concerning the $%
K\overline{K}-\pi \pi$ scattering, it remains in the same energy range as in 
$\phi\rightarrow \pi\pi\gamma$ decays and we can safely use the amplitudes
calculated in unitarized chiral perturbation theory which contains naturally
the scalar poles.

The high virtuality of the exchanged photon makes probable the excitation of
higher mass hadronic states. The quark structure of the $\phi$ suggests that
the $K^{\ast}K$ intermediate state can also give important contributions to $%
e^{+}e^{-}\rightarrow\phi\pi\pi$ via the production of virtual $K^{\ast }%
\overline{K}$ , with the virtual $K^{\ast}$ decaying into a $\phi$ $K$ and
the final rescattering of kaons into pions. In this concern it is worth
mentioning that experimental data on $e^{+}e^{-}\longrightarrow K^{0}K^{\pm
}\pi^{\mp}$ at $\sqrt{s}=$ $1400-2180\ MeV$ shows that this reaction is
dominated by intermediate neutral $K^{\ast0}K^{0}$ production with the $%
K^{\ast0}$ decaying into $K^{\pm}\pi^{\mp}$\cite{BiselloCS}, hence there is
a sizable coupling of a virtual photon to the $K^{\ast}K$ system at the
mentioned energies. The proper description of this mechanism requires the
knowledge of the transition $K^{\ast}K$ electromagnetic form factor but,
again, it can be extracted from experimental data on $e^{+}e^{-}%
\longrightarrow K^{0}K^{\pm}\pi^{\mp}$ which shows that, in addition to the
contributions from the exchange of lowest lying vectors, this form factor
receives also contributions from the exchange of $\phi^{\prime}$ and $%
\rho^{\prime}$. Remarkably there is no evidence for contributions coming
from the exchange of $\omega^{\prime}$ to this form factor.

In this paper we study the above mentioned mechanisms for $%
e^{+}e^{-}\rightarrow\phi\pi\pi$ for the dipion system in an isoscalar $s-$%
wave. The paper is organized as follows: In section II we calculate the $%
\gamma^{\ast}\phi\pi\pi$ vertex function using $U\chi PT$. In
section III we calculate intermediate vector meson contributions using 
$U\chi PT$ supplemented with the anomalous term describing $VVP$ interactions
and incorporate contributions from heavy mesons to the $K^{\ast}K$
transition form factor. In section IV we analyze the different contributions
and our summary and conclusions are given in section V.

\section{Unitarized $\protect\chi PT$ predictions for $e^{+}e^{-}\rightarrow 
\protect\phi\left( \protect\pi\protect\pi\right) _{I=J=0}$.}

\begin{figure}[ptb]
\begin{center}
\includegraphics[
natheight=11.170200in,
natwidth=9.180100in,
height=5.50in,
width=4.75in
]{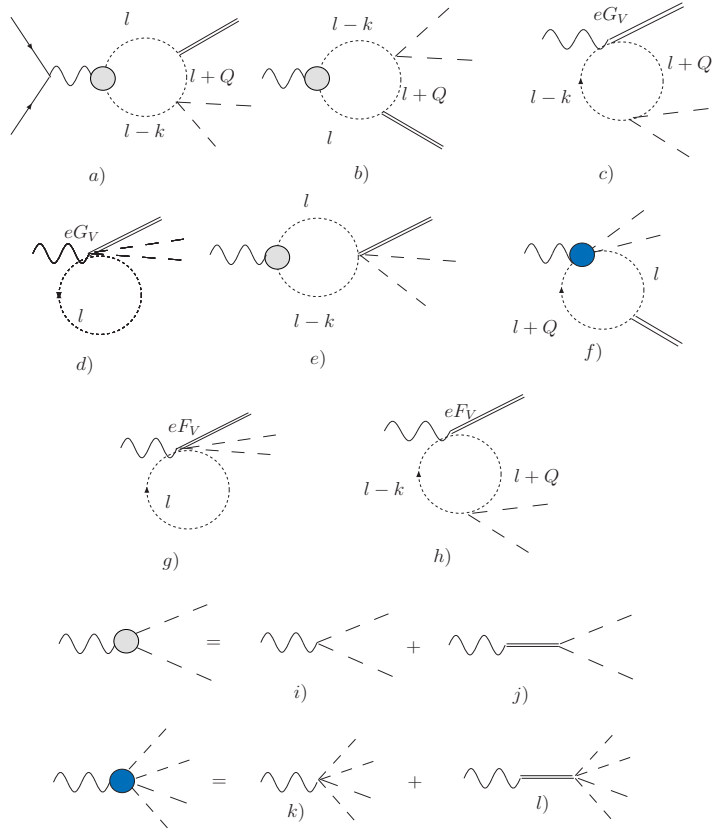}
\end{center}
\caption{Feynman diagrams for $e^{+}e^{-}\rightarrow\protect\phi\protect\pi%
\protect\pi$ in $R\protect\chi PT$.}
\label{FD}
\end{figure}

Following \cite{MHOT}, the process $e^{+}e^{-}\rightarrow\phi\pi\pi$ is
induced at one loop level by the kaon loops. In the calculations the
vertices are borrowed from Resonance Chiral Perturbation Theory ($R\chi PT$) 
\cite{EGPR}. We follow the conventions in \cite{EGPR} and the relevant
interactions in their notation are%
\begin{eqnarray}
\mathcal{L} & =&\mathcal{L}^{(2)}+\mathcal{L}^{(F)}+\mathcal{L}^{(G)}
\label{lag} \\
\mathcal{L}^{(2)} & =&\frac{1}{4}f^{2}tr\left( \left( D_{\mu}U\right)
^{\dagger}D^{\mu}U+\chi U^{\dagger}+\chi^{\dagger}U\right)  \label{L2} \\
\mathcal{L}^{(F)} & =&\frac{F_{V}}{2\sqrt{2}}tr(V_{\mu\nu}f_{+}^{\mu\nu })
\label{LF} \\
\mathcal{L}^{(G)} & =&\frac{iG_{V}}{\sqrt{2}}tr(V_{\mu\nu}u^{\mu}u^{\nu}),
\label{LG}
\end{eqnarray}
where 
\begin{eqnarray}
u_{\mu} & =&iu^{\dagger}\ D_{\mu}U\ u^{\dagger},\qquad U=u^{2},\qquad u=e^{-%
\frac{i}{\sqrt{2}}\frac{\Phi}{f}},\qquad\Phi=\frac{1}{\sqrt{2}}%
\lambda_{i}\varphi_{i}  \label{umu} \\
f_{+}^{\mu\nu} & =&u\ F_{L}^{\mu\nu}u^{\dagger}+u^{\dagger}\ F_{R}^{\mu\nu }\
u,\qquad D_{\mu}U=\partial_{\mu}U-i\left[ v_{\mu},U\right] .
\end{eqnarray}
We introduce the photon field through $v_{\mu}=eQA_{\mu}$\ and $%
F_{L}^{\mu\nu }=F_{R}^{\mu\nu}=eQF^{\mu\nu}$ ($e>0$) where $F^{\mu\nu}\ $%
denotes the electromagnetic strenght tensor. For further details in the
notation we refer the reader to Ref. \cite{EGPR}. The relevant diagrams are
shown in Fig. (\ref{FD}), where for simplicity a shaded circle and a dark
circle account for the diagrams $i)$ plus $j)$ and $k)$ plus $l)$
respectively, which differentiate the direct photon coupling from the
coupling through an intermediate vector meson. We will address the
corresponding diagrams as $a)$, $b)$, when we have the direct photon
coupling and $a^{\prime})$, $b^{\prime})$, when the coupling goes through
the exchange of a vector meson. The kaon form factor at lowest order
contains the exchange of vector mesons in diagrams $a^{\prime}),b^{\prime})$
which in $R\chi PT$ are intrinsically gauge invariant.

One interesting feature of the use of meson-meson chiral amplitudes is that
in the different processes one can factorize the amplitude on-shell inside
the loops. This is the case in the construction of the unitary meson-meson
amplitudes where the factorization can be seen as a consequence of the
reabsorption of the off-shell terms into renormalization of elementary
couplings \cite{OO}, or using the $N/D$ method of unitarization that relies
upon the imaginary part of the amplitudes which involves the on-shell part 
\cite{ND}. These two methods have been generalized to the case of
meson-baryon interaction in \cite{Ramos} and \cite{Ulf} respectively. More
concretely, for the case close to ours in $\phi\rightarrow K^{0}\overline{%
K^{0}}\gamma$ it was demonstrated, using arguments of gauge invariance, that
only the on-shell part of the meson-meson amplitudes was needed inside the
loops \cite{Oller}. Explicit cancellation of the off-shell terms can be seen
in our formalism and we only sketch the derivation since there are basic
principles that tells us this factorization should always be possible. The
reason is that the off-shell part of the meson-meson amplitude is unphysical
and can be changed with a unitary transformation of the fields, that,
however, should not change the physical amplitudes. Technically the
cancellations in our formalism go as follows. As discussed in \cite{OO,OOPel}%
, to lowest order in the chiral expansion the $K\overline{K}-\pi\pi$
amplitude (denoted by $\widetilde{V}_{K\pi}^{0}$) for arbitrary values of
the particle momenta $p_{i}$ has the form%
\begin{equation}
\widetilde{V}_{K\pi}^{0}=V_{K\pi}^{0}+\beta\sum_{i}(p_{i}^{2}-m_{i}^{2}),
\label{fact}
\end{equation}
where $V_{K\pi}^{0}$ denotes the on-shell amplitude. In the following we use
the convention that all external particle momenta of the $%
\gamma^{\ast}(k)\phi(Q)\pi(p)\pi(p^{\prime})$ vertex function flow into the
vertices and will change this direction only in the numerical results.
Considering the off-shell part of the meson-meson interaction in diagrams $%
a) $, $b)$, associated to the line of momentum $l-k$ cancels the
corresponding meson propagator and generates a topological structure like
the one of diagram $f)$. On the other hand, diagram $f)$ is a genuine
diagram that can be calculated by using the Lagrangian $\mathcal{L}^{2}$ of
Eq. (\ref{L2}) expanded to four mesons. When this is done one finds an exact
cancellation of the off-shell terms against diagram $f)$. On the other hand,
there are similar cancellations between the off-shell part of the
meson-meson amplitude associated to the line with momentum $l+Q$ in diagrams 
$a)$, $b)$ and $c)$ with the genuine contributions in diagrams $d)$ and $e)$%
. A remnant contribution appears after the cancellations, which vanishes for
real photons and involves derivatives in the vector fields. Exact
cancellation of this part would require the introduction of counterterm
Lagrangians involving derivatives of $V^{\mu\nu}$ and $f^{\mu\nu}_{+}$ and
such Lagrangians are sometimes used for this purpose \cite{Pich}. Finally
the off-shell part of diagram $h)$ which involve charged kaons only cancels
exactly diagram $g)$ with charged kaons in the loops. Remaining tadpole
contributions from neutral kaons can be cancelled by appropriate
counterterms. In summary, all one has to do is to evaluate the diagrams $a)$%
, $b)$, $c)$, and $h)$ with the meson-meson amplitudes factorized on-shell
and omiting the rest of diagrams. This lowest order amplitude is iterated 
in the coupled channel framework used in \cite{OO,OOPel} to obtain 
the unitarized $K\bar{K}\to\pi\pi$ amplitudes which contain the scalar poles. 
In a coming section we will study the
contributions of loops involving vector meson propagators. In this case we
do not have enough information on the higher order Lagrangians to explicitly
show the cancellations but we shall equally assume that the meson-meson
amplitude can be factorized on-shell out of the loops.

Let us start with the simplest diagrams $a)$, $b)$ with charged kaons in the
loops, point-like $K^{+}K^{-}\gamma$ interaction and charged pions in the
final state. A straightforward calculation yields%
\begin{equation}
-i\mathcal{M}_{K^{+}}^{a+b}=-\frac{e^{2}\sqrt{2}G_{V}}{f^{2}}\frac{t_{K\pi
}^{0}}{\sqrt{3}}\frac{L^{\mu}}{k^{2}}T_{\mu\nu}Q_{\alpha}\eta^{\alpha\nu } 
\end{equation}
where $k^{2}=(p^{+}+p^{-})^{2}$, $L^{\mu}\equiv\overline{v}%
(p^{+})\gamma^{\mu }u(p^{-})$, $Q$ denotes the momentum of the $\phi$ and $%
\eta^{\alpha\nu}$ denotes the polarization tensor of the anti-symmetric
field $\phi_{\mu\nu}$ used to describe the $\phi$ meson. The on-shell
unitarized amplitude for isoscalar $s$-wave $K\overline{K}-\pi\pi$
scattering is denoted as $t_{K\pi }^{0}$and it is related to the physical $%
t_{K^{+}\pi^{+}}$ amplitude as $t_{K^{+}\pi^{+}}=\frac{t_{K\pi}^{0}}{\sqrt{3}%
}$. It factorizes on-shell out of the loop tensor integral given by 
\begin{equation}
T_{\mu\nu}=i\int\frac{d^{4}l}{(2\pi)^{4}}\frac{2(2l-k)_{\mu}l_{\nu}}{%
\square\left( l+Q\right) \square(l)\square\left( l-k\right) },
\end{equation}
with $\square(l)\equiv l^{2}-m_{K}^{2}+i\varepsilon$.

The "seagull" diagram $c)$ yields%
\begin{equation}
-i\mathcal{M}_{K^{+}}^{c}=\frac{e^{2}\sqrt{2}G_{V}}{f^{2}}\frac{t_{K\pi}^{0}%
}{\sqrt{3}}\frac{L^{\mu}}{k^{2}}G_{K}(m_{\pi\pi}^{2})g_{\mu\nu}\left(
Q+k\right) _{\alpha}\eta^{\alpha\nu},  
\end{equation}
\ where $m_{\pi\pi}^{2}=(Q+k)^{2}$ and $G_{K}$ denotes the loop integral%
\begin{equation}
G_{K}(p^{2})\equiv\int\frac{d^{4}l}{(2\pi)^{4}}\frac{i}{\square\left(
l\right) \square\left( l+p\right) }.  \label{G}
\end{equation}

Using dimensional regularization we get  
\begin{equation}
G_{K}(m_{\pi\pi}^{2})=\mu^{2\varepsilon}\int\frac{d^{d}l}{(2\pi)^{d}}\frac {i%
}{\square\left( l+Q\right) \square\left( l-k\right) }=\frac{1}{\left(
4\pi\right) ^{2}}\left( a(\mu)+\log\frac{m_{K}^{2}}{\mu^{2}}+I_{G}(m_{\pi
\pi}^{2})\right)  \label{GKreg}
\end{equation}
with%
\begin{equation}
I_{G}=\int_{0}^{1}dx\log\left( 1-\frac{m_{\pi\pi}^{2}}{m_{K}^{2}}%
x(1-x)-i\varepsilon\right) =-2+\sigma\log\frac{\sigma+1}{\sigma-1},
\end{equation}
where $\sigma(m_{\pi\pi}^{2})=\sqrt{1-\frac{4m_{K}^{2}}{m_{\pi\pi}^{2}}}$.
The substraction constant has been fixed in Ref. \cite{OOPel} to $%
a(\mu_{0})=1$ for $\mu_{0}=1.2$ $GeV$ matching the cutoff regularized
integral for a cutoff $\Lambda=1$ $GeV$. It is related at different scales
as $a(\mu)=a(\mu _{0})+\log\frac{\mu^{2}}{\mu_{0}^{2}}$ in such a way that
the loop function is scale independent.

There is no direct coupling of the photon to neutral kaons and adding up all
contributions we obtain 
\begin{equation}
-i\mathcal{M}_{K}^{a+b+c}=-\frac{e^{2}\sqrt{2}G_{V}}{f^{2}}\frac{t_{K\pi}^{0}%
}{\sqrt{3}}\frac{L^{\mu}}{k^{2}}\left(
T_{\mu\nu}^{abc}Q_{\alpha}-G_{K}(m_{\pi\pi}^{2})g_{\mu\nu}k_{\alpha}\right)
\eta^{\alpha\nu}  \label{mabc}
\end{equation}
where%
\begin{equation}
T_{\mu\nu}^{abc}=T_{\mu\nu}-G_{K}(m_{\pi\pi}^{2})g_{\mu\nu}.  \label{tmunuq}
\end{equation}
Notice that in diagrams $a)$, $b)$, $c)$ pions appear only through $t_{K\pi
}^{0}$. Since the $K\overline{K}-\pi\pi$ amplitude with neutral pions
satisfy $t_{K^{+}\pi^{0}}=\frac{t_{K\pi}^{0}}{\sqrt{3}}$ the amplitude for $%
e^{+}e^{-}\rightarrow\phi\pi\pi$ with neutral pions in the final state is
also given by Eq. (\ref{mabc}).

Let us now consider diagrams $a^{\prime})$, $b^{\prime})$ with charged kaons
in the loops and charged pions in the final state. These diagrams involve
the propagation of vector particles. The propagator for a vector meson in
the tensor formalism is given by%
\begin{equation}
\Pi_{\alpha\beta\mu\nu}(p)=\frac{i\Delta_{\alpha\beta\mu\nu}(p)}{%
p^{2}-M_{V}^{2}+i\varepsilon}
\end{equation}
where%
\begin{equation}
\Delta_{\mu\nu\rho\sigma}(p)=\frac{1}{M_{V}^{2}}\left[ \left(
p^{2}-M_{V}^{2}\right) g_{\mu\rho}g_{\nu\sigma}-g_{\mu\rho}p_{\nu}p_{\sigma
}+g_{\mu\sigma}p_{\nu}p_{\rho}-\left( \mu\leftrightarrow\nu\right) \right].
\end{equation}
This tensor is anti-symmetric under the exchange $\mu\leftrightarrow\nu$ or $%
\rho\leftrightarrow\sigma$, symmetric under the exchange $\mu\nu
\leftrightarrow\rho\sigma$ and satisfy 
\begin{equation}
p^{\mu}\Delta_{\mu\nu\rho\sigma}(p)=g_{\nu\rho}p_{\sigma}-g_{\nu\sigma}p_{%
\rho},\qquad\Delta_{\mu\nu\rho\sigma}(p)p^{\sigma}=g_{\nu\rho}p_{\mu
}-g_{\mu\sigma}p_{\nu}.  \label{simple}
\end{equation}
The Lagrangian in Eq. (\ref{lag}) yield the following vertices for the $%
\gamma(k,\mu)V(k,\alpha\beta)$ and $V(Q,\alpha\beta)P(p)P^{\prime}(p^{\prime
})$ interactions 
\begin{equation}
\Gamma_{\mu\alpha\beta}^{\gamma V}=\frac{eF_{V}}{3}k_{\alpha}g_{\mu%
\beta}C_{V},\qquad\Gamma_{\alpha\beta}^{VPP^{\prime}}=-\frac{\sqrt{2}%
G_{V}C_{VPP^{\prime}}}{f^{2}}p_{\alpha}p_{\beta}^{\prime},
\end{equation}
with the $SU(3)$ factors given by%
\begin{eqnarray}
C_{\phi} &=&-\sqrt{2},\ \quad C_{\omega}=1,\ \quad C_{\rho}=3;  \label{cv} \\
C_{\phi K^{+}K^{-}} &=&C_{\phi K^{0}\overline{K}^{0}}=1,\quad C_{\omega
K^{+}K^{-}}=C_{\omega K^{0}\overline{K}^{0}}=-\frac{1}{\sqrt{2}},\quad
C_{\rho K^{+}K^{-}}=-\frac{1}{\sqrt{2}},\ C_{\rho K^{0}\overline{K}^{0}}=%
\frac {1}{\sqrt{2}}.  \label{ccv}
\end{eqnarray}
The amplitude for diagrams $a^{\prime}),b^{\prime})$ whith charged kaons in
the loops is%
\begin{equation}
-i\mathcal{M}_{K^{+}}^{a^{\prime}+b^{\prime}}=\frac{\sqrt{2}e^{2}G_{V}}{f^{2}%
}\frac{L^{\mu}}{k^{2}}\frac{t_{K\pi}^{0}}{\sqrt{3}}\widetilde{F}%
_{K^{+}}(k^{2})T_{\mu\alpha\nu}^{a^{\prime}b^{\prime}}\eta^{\alpha\nu}
\label{mapbp0}
\end{equation}
where $\widetilde{F}_{K^{+}}(k^{2})$ stands for the vector meson
contributions to the charged kaon form factor 
\begin{equation}
\widetilde{F}_{K^{+}}(k^{2})=\frac{1}{2}\sum_{V=\rho,\phi,\omega}\frac{F_{V}%
}{3}\frac{\sqrt{2}G_{V}C_{V}C_{VK^{+}K^{-}}}{f^{2}}\frac{k^{2}}{%
k^{2}-M_{V}^{2}}=\frac{G_{V}F_{V}}{2f^{2}}\left( \frac{k^{2}}{%
m_{\rho}^{2}-k^{2}}+\frac{1}{3}\frac{k^{2}}{m_{\omega}^{2}-k^{2}}+\frac{2}{3}%
\frac{k^{2}}{m_{\phi}^{2}-k^{2}}\right) .  \label{ftilde}
\end{equation}
and the loop tensor integral is given by 
\begin{equation}
T_{\mu\alpha\nu}^{a^{\prime}b^{\prime}}=\frac{1}{k^{2}}k^{\sigma}\Delta_{%
\sigma\mu\gamma\delta}(k)i\int\frac{d^{4}l}{(2\pi)^{4}}\frac {%
4(l-k)^{\gamma}l^{\delta}l_{\alpha}(l+Q)_{\nu}}{\square_{K}(l)\square
_{K}\left( l+Q\right) \square_{K}\left( l-k\right) }.
\end{equation}
This is an explicitly gauge invariant tensor due to the anti-symmetry of $%
\Delta_{\sigma\mu\gamma\delta}(k)$ under $\sigma\leftrightarrow\mu$. Using $%
\Delta_{\sigma\mu\gamma\delta}(k)k^{\gamma}k^{\delta}=0$ and $\eta^{\alpha
\nu}=-\eta^{\nu\alpha}$ it can be rewritten to 
\begin{equation}
T_{\mu\alpha\nu}^{a^{\prime}b^{\prime}}=-\left( T_{\mu\nu}^{abc}+\frac {%
G_{K}(m_{\pi\pi}^{2})}{k^{2}}\left( k^{2}g_{\mu\nu}-k_{\mu}k_{\nu}\right)
\right) Q_{\alpha}.
\end{equation}
The amplitude for diagrams $a^{\prime}),b^{\prime})$ can in turn be
rewritten as%
\begin{equation}
-i\mathcal{M}_{K^{+}}^{a^{\prime}+b^{\prime}}=\frac{-e^{2}\sqrt{2}G_{V}}{%
f^{2}}\frac{t_{K\pi}^{0}}{\sqrt{3}}\frac{L^{\mu}}{k^{2}}\widetilde{F}%
_{K^{+}}(k^{2})\left( T_{\mu\nu}^{abc}+\frac{G_{K}(m_{\pi\pi}^{2})}{k^{2}}%
\left( k^{2}g_{\mu\nu}-k_{\mu}k_{\nu}\right) \right)
Q_{\alpha}\eta^{\alpha\nu}.  \label{mapbpkp}
\end{equation}
There are also contributions of neutral kaons in the loops. The calculation
of these contributions is similar to the charged kaon loops due to the
related $SU(3)$ factors in Eq. (\ref{ccv}). The only difference comes from
the sign of the $\rho$ factors in Eq. (\ref{ccv}) which changes from the
charged to the neutral case. The total amplitude is obtained from Eq.(\ref%
{mapbpkp}) just replacing $\widetilde{F}_{K^{+}}$ by $\widetilde{F}_{K^{+}}+$
$\widetilde {F}_{K^{0}}$ where the intermediate $\rho$ contributions cancel.
Including neutral and charged kaon contribution we obtain 
\begin{equation}
-i\mathcal{M}_{K}^{a^{\prime}+b^{\prime}}=\frac{-e^{2}\sqrt{2}G_{V}}{f^{2}}%
\frac{t_{K\pi}^{0}}{\sqrt{3}}\frac{L^{\mu}}{k^{2}}\widetilde{F}%
_{iso}(k^{2})\left( T_{\mu\nu}^{abc}+\frac{G_{K}(m_{\pi\pi}^{2})}{k^{2}}%
\left( k^{2}g_{\mu\nu}-k_{\mu}k_{\nu}\right) \right)
Q_{\alpha}\eta^{\alpha\nu},  \label{mapbp}
\end{equation}
with%
\begin{equation}
\widetilde{F}_{iso}(k^{2})=\widetilde{F}_{K^{+}}(k^{2})+\widetilde{F}%
_{K^{0}}(k^{2})=\frac{F_{V}G_{V}}{3f^{2}}\left( \frac{k^{2}}{%
m_{\omega}^{2}-k^{2}}+\frac{2k^{2}}{m_{\phi}^{2}-k^{2}}\right) .
\label{Fiso}
\end{equation}
For neutral pions in the final state we obtain the same result due to
relations $t_{K^{+}\pi^{0}}=\frac{t_{K\pi}^{0}}{\sqrt{3}}$ and $%
t_{K^{0}\pi^{0}}=\frac{t_{K\pi}^{0}}{\sqrt{3}}$.

The calculation of diagram $h)$ requires to work out the $\gamma(k,\mu
)\phi(Q,\alpha\nu)K(p)\overline{K}(p^{\prime})$ vertex contained in $%
\mathcal{L}^{F}$ in Eq. (\ref{LF}). For neutral kaons this vertex vanishes
and for charged kaons we obtain%
\begin{equation}
\Gamma_{\mu\alpha\nu}=\frac{eF_{V}}{\sqrt{2}f^{2}}g_{\mu\nu}k_{\alpha}.
\end{equation}
The amplitude for diagram $h)$ is%
\begin{equation}
-i\mathcal{M}^{h}=-\frac{e^{2}F_{V}}{\sqrt{2}f^{2}}\frac{L^{\mu}}{k^{2}}%
\frac{t_{K\pi}^{0}}{\sqrt{3}}G_{K}(m_{\pi\pi}^{2})g_{\mu\nu}k_{\alpha}\eta^{%
\alpha\nu}.  \label{mh}
\end{equation}
Adding up contributions of all diagrams in Eqs. (\ref{mabc},\ref{mapbp},\ref%
{mh}) we obtain the kaon loop contributions for both final pion charge
states as%
\begin{align}
-i\mathcal{M}_{K} & =-\frac{e^{2}\sqrt{2}G_{V}}{f^{2}}\frac{t_{K\pi}^{0}}{%
\sqrt{3}}\frac{L^{\mu}}{k^{2}}\left( F_{VMD}^{0}(k^{2})T_{\mu\nu}^{abc}+%
\widetilde{F}_{iso}(k^{2})\frac{G_{K}(m_{\pi\pi}^{2})}{k^{2}}\left(
k^{2}g_{\mu\nu}-k_{\mu}k_{\nu}\right) \right) Q_{\alpha}\eta^{\alpha\nu }
\label{mk} \\
& +\frac{e^{2}\sqrt{2}}{f^{2}}(G_{V}-\frac{F_{V}}{2})\frac{t_{K\pi}^{0}}{%
\sqrt{3}}\frac{L^{\mu}}{k^{2}}G_{K}(m_{\pi\pi}^{2})g_{\mu\nu}k_{\alpha}%
\eta^{\alpha\nu},  \notag
\end{align}
where%
\begin{equation}
F_{VMD}^{0}(k^{2})=1+\widetilde{F}_{iso}(k^{2})=1+\frac{F_{V}G_{V}}{3f^{2}}%
\left( \frac{k^{2}}{m_{\omega}^{2}-k^{2}}+\frac{2k^{2}}{m_{\phi}^{2}-k^{2}}%
\right) .
\end{equation}
accounts for the lowest order terms of the kaon isoscalar form factor (the
sum of the charged and neutral kaon form factors) in $R\chi PT$ \cite{OOP}
which is valid at low photon virtualities. Notice that the second term in
Eq. (\ref{mk}) contains only the vector meson contributions to the kaon form
factor but the constant term due to the electric charge is missing. This
term should come from Lagrangians with higher derivatives of the fields (
specifically from the term $\partial^{\alpha}V_{\alpha\nu}
\partial_{\mu}f^{\mu\nu}_{+}$ ) which is absent in our basic interactions in
Eq. (\ref{lag}). We will assume in the following that the constant term due
to the charge is provided by such missing interactions and, hence, write $%
F_{VMD}^{0} $ instead of $\widetilde{F}_{iso}$ in the second term of Eq. (%
\ref{mk}).

The high virtualities involved in our process requires to work out the
complete $\gamma K\overline{K}$ vertex functions. The calculation of these
vertex functions has been done in the context of $U\chi PT$ in Ref. \cite%
{OOP}. We use this result and replace in the following the leading order
terms so far obtained, $F_{VMD}^{0}(k^{2})$, by the full isoscalar form
factor $F_{K}^{0}(k^{2})=F_{K^{+}}(k^{2})+F_{K^{0}}(k^{2})$.

The evaluation of Eq. (\ref{mk}) requires to work out the loop tensor $%
T_{\mu \nu }^{abc}$. It can be easily shown that $T_{\mu \nu }^{abc}$ is
finite and gauge invariant. The most general form of this tensor is 
\begin{equation}
T_{\mu \nu }^{abc}=a\ g_{\mu \nu }+b\ Q_{\mu }Q_{\nu }+c\ Q_{\mu }k_{\nu
}+d\ k_{\mu }Q_{\nu }+e\ k_{\mu }k_{\nu }
\end{equation}%
where $\ a,b,c,d,e$ are form factors. Gauge invariance requires 
\begin{equation}
k^{\mu }T_{\mu \nu }^{abc}=\left( a+ck\cdot Q+ek^{2}\right) k_{\nu
}+(bk\cdot Q+dk^{2})Q_{\nu }=0,
\end{equation}%
imposing the following relations among the form factors%
\begin{equation}
a=-c\ k\cdot Q-e\ k^{2},\qquad b\ k\cdot Q=-d\ k^{2},
\end{equation}%
thus $T_{\mu \nu }^{abc}$ has the following explicitly gauge invariant form 
\begin{equation}
T_{\mu \nu }^{abc}=-c(Q\cdot k\ g_{\mu \nu }-Q_{\mu }k_{\nu })-\frac{d}{%
k\cdot Q}\ (k^{2}Q_{\mu }-k\cdot Qk_{\mu })Q_{\nu }-e\ (k^{2}g_{\mu \nu
}-k_{\mu }k_{\nu }).  \label{git}
\end{equation}%
The second term vanishes upon contraction with $Q_{\alpha }\eta ^{\alpha \nu
}$ and we are left only with two form factors 
\begin{equation}
T_{\mu \nu }^{abc}=-c(Q\cdot k\ g_{\mu \nu }-Q_{\mu }k_{\nu })-e\
(k^{2}g_{\mu \nu }-k_{\mu }k_{\nu }).  \label{tmunugen}
\end{equation}%
A straightforward calculation using conventional Feynman parametrization
yields 
\begin{equation}
c=-\frac{1}{4\pi ^{2}m_{K}^{2}}I_{P},\qquad e=-\frac{1}{4\pi ^{2}m_{K}^{2}}%
J_{P}.  \label{ce}
\end{equation}%
where%
\begin{eqnarray}
I_{P}& =\int_{0}^{1}dx\int_{0}^{x}dy\frac{y(1-x)}{1-\frac{Q^{2}}{m_{K}^{2}}%
x(1-x)-\frac{2Q\cdot k}{m_{K}^{2}}(1-x)y-\frac{k^{2}}{m_{K}^{2}}%
y(1-y)-i\varepsilon }  \label{IAB} \\
J_{P}& =\frac{1}{2}\int_{0}^{1}dx\int_{0}^{x}dy\frac{y(1-2y)}{1-\frac{Q^{2}}{%
m_{K}^{2}}x(1-x)-\frac{2Q\cdot k}{m_{K}^{2}}(1-x)y-\frac{k^{2}}{m_{K}^{2}}%
y(1-y)-i\varepsilon }.  \label{JAB}
\end{eqnarray}%
In terms of the $\ a$ and $e$ form factors we get the amplitude for kaon
loops contribution to $e^{+}e^{-}\rightarrow \gamma ^{\ast }\rightarrow \phi
\lbrack \pi \pi ]_{I,J=0}$ as%
\begin{align}
-i\mathcal{M}_{P}& =\frac{e^{2}\sqrt{2}G_{V}}{f^{2}}\frac{t_{K\pi }^{0}}{%
\sqrt{3}}\frac{L^{\mu }}{k^{2}}F_{K}^{0}(k^{2})\left[ c(Q\cdot k\ g_{\mu \nu
}-Q_{\mu }k_{\nu })+\left( e-\frac{G_{K}}{k^{2}}\right) (k^{2}g_{\mu \nu
}-k_{\mu }k_{\nu })\right] Q_{\alpha }\eta ^{\alpha \nu }  \label{mp} \\
& +\frac{e^{2}\sqrt{2}}{f^{2}}(G_{V}-\frac{F_{V}}{2})\frac{t_{K\pi }^{0}}{%
\sqrt{3}}\frac{L^{\mu }}{k^{2}}G_{K}(m_{\pi \pi }^{2})g_{\mu \nu }k_{\alpha
}\eta ^{\alpha \nu }.  \notag
\end{align}%
The vertex function for $\gamma ^{\ast }(k)\phi (Q,\alpha \nu )[\pi (q)\pi
(q^{\prime })]_{I,J=0}$ is straightforwardly obtained just removing the
factor $-\frac{eL^{\mu }}{k^{2}}$ and it is worthy to analyze our results in
terms of this vertex function. Notice that in addition to the terms
associated to the full kaon form factors we get a contact term which
survives in the real photon limit and has been already noticed in the
studies of radiative $\phi $ decays \cite{MHOT}. The combination $G_{V}-%
\frac{F_{V}}{2}$ is small and it vanishes in the context of Vector Meson
Dominance \cite{GVFV}. We will keep this term and discuss below its impact 
on the cross
section but we must be clear from the beggining that it can not be taken
seriously at high photon virtualities without its dressing by a form factor.

Tensor and vector fields are related as $\partial ^{\mu }V_{\mu \nu
}=M_{V}V_{\nu }$ and for an on-shell $\phi $ it is convenient to rewrite Eq.
(\ref{mp}) in terms of the conventional polarization vector related to the
polarization tensor as $\eta ^{\alpha \nu }(Q)=\frac{i}{M_{\phi }}(Q^{\alpha
}\eta ^{\nu }-Q^{\nu }\eta ^{\alpha })$ in such a way that 
\begin{equation}
Q_{\alpha }\eta ^{\alpha \nu }(Q)=iM_{\phi }\eta ^{\nu }(Q),\qquad g_{\mu
\nu }k_{\alpha }\eta ^{\alpha \nu }(Q)=\frac{i}{M_{\phi }}(Q\cdot kg_{\mu
\nu }-Q_{\mu }k_{\nu })\eta ^{\nu }.  \label{keta}
\end{equation}%
Using these relations we get%
\begin{align}
-i\mathcal{M}_{P}& =\frac{ie^{2}\sqrt{2}M_{\phi }}{f^{2}}\frac{t_{K\pi }^{0}%
}{\sqrt{3}}\frac{L^{\mu }}{k^{2}}\left[ \left( G_{V}F_{K}^{0}(k^{2})c+(G_{V}-%
\frac{F_{V}}{2})\frac{G_{K}(m_{\pi \pi }^{2})}{M_{\phi }^{2}}\right) (Q\cdot
k\ g_{\mu \nu }-Q_{\mu }k_{\nu })\right. \\
& \left. +G_{V}F_{K}^{0}(k^{2})\left( e-\frac{G_{K}}{k^{2}}\right)
(k^{2}g_{\mu \nu }-k_{\mu }k_{\nu })\right] \eta ^{\nu }  \notag
\end{align}%
Using now Eqs. (\ref{ce}) we obtain
\begin{equation}
-i\mathcal{M}_{P}=\frac{-ie^{2}}{2\pi ^{2}m_{K}^{2}}\frac{t_{K\pi }^{0}}{%
\sqrt{3}}\frac{L^{\mu }}{k^{2}}\left[ A_{P}\ L_{\mu \nu }^{(1)}+B_{P}L_{\mu
\nu }^{(2)}\right] \eta ^{\nu }
\end{equation}%
with the Lorentz structures%
\begin{equation}
L_{\mu \nu }^{(1)}\equiv Q\cdot kg_{\mu \nu }-Q_{\mu }k_{\nu },\qquad L_{\mu
\nu }^{(2)}=k^{2}g_{\mu \nu }-k_{\mu }k_{\nu },  \label{L12}
\end{equation}%
and 
\begin{eqnarray}
A_{P}& =&\frac{\sqrt{2}M_{\phi }}{2f^{2}}\left(
G_{V}F_{K}^{0}(k^{2})I_{P}-(G_{V}-\frac{F_{V}}{2})\frac{m_{K}^{2}}{4M_{\phi
}^{2}}g_{K}(m_{\pi \pi }^{2})\right) ,  \label{AP} \\
B_{P}& =&\frac{\sqrt{2}M_{\phi }G_{V}}{2f^{2}}F_{K}^{0}(k^{2})\left( J_{P}+%
\frac{m_{K}^{2}}{4k^{2}}g_{K}\right) .\label{BP}
\end{eqnarray}%
where we defined $g_{K}(p^{2})\equiv (4\pi )^{2}G_{K}(p^{2})$.

\section{Contributions from vectors in the loops}

\begin{figure}[ptb]
\begin{center}
\includegraphics[
natheight=11.830200in,
natwidth=8.919900in,
height=4.0in,
width=3.5in
]{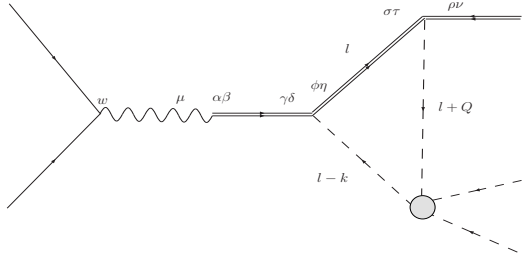}
\end{center}
\caption{Feynman diagram for $e^{+}e^{-}\to K^{*}\bar{K}\to \protect\phi K 
\bar{K}\to \protect\phi \protect\pi\protect\pi$. }
\label{FDV}
\end{figure}

The process $e^{+}(p^{+})e^{-}(p^{-})\rightarrow\phi(Q,\eta)\ \pi (p)\
\pi(p^{\prime})$\ can also proceed through $e^{+}(p^{+})e^{-}(p^{-})%
\rightarrow K^{\ast}(p)\ \overline{K}(p^{\prime})\rightarrow\phi (Q,\eta)\
K(p)\ \overline{K}(p^{\prime})$\ with the kaons rescattering to a pion pair
as shown in Fig. (\ref{FDV}). 
The $VV^{\prime}P$ interaction is dictated by the anomalous Lagrangian which
we rewrite in terms of the tensor field as 
\begin{equation}
\mathcal{L}_{anom}=\frac{G}{\sqrt{2}}\epsilon_{\mu\nu\alpha\beta}tr(%
\partial^{\mu}V^{\nu}\partial^{\alpha}V^{\beta}\Phi)=\frac{G_{T}}{4\sqrt {2}}%
\epsilon_{\mu\nu\alpha\beta}tr(V^{\mu\nu}V^{\alpha\beta}\Phi).
\end{equation}
with $G_{T}=M_{V}M_{V^{\prime}}G$. The required vertex for $V(k,\mu
,\nu)V^{\prime}(q,\alpha,\beta)P$ is 
\begin{equation}
\Gamma_{\mu\nu\alpha\beta}(k,q)=\frac{iG_{T}C_{VV^{\prime}P}}{4\sqrt{2}}%
\epsilon_{\mu\nu\alpha\beta}
\end{equation}
with the SU(3) factors given by 
\begin{equation}
C_{\phi K^{\ast+}K^{-}}=C_{\phi K^{\ast0}K^{0}}=1,\qquad C_{\rho K^{\ast
+}K^{-}}=-C_{\rho K^{\ast0}K^{0}}=C_{\omega K^{\ast+}K^{-}}=C_{\omega
K^{\ast0}K^{0}}=\frac{1}{\sqrt{2}}.
\end{equation}
The amplitude from the diagram in Fig. (\ref{FDV}) gets contributions from $%
K^{\ast+}K^{-}$and $K^{\ast-}K^{+}$ in the loops plus $K^{\ast0}\overline{%
K^{0}}$and $\overline{K^{\ast0}}K^{0}$. The first two contributions can be
summed to%
\begin{equation}
-i\mathcal{M}_{+}=-2e^{2}F_{K^{\ast+}K^{-}}^{lo}(k^{2})\frac{G_{T}}{\sqrt{2}}%
\left( \frac{M_{K^{\ast}}}{16}\right) \frac{t_{K\pi}^{0}}{\sqrt{3}}\frac{%
L^{\mu}}{k^{2}}T_{\mu\alpha\nu}\eta^{\alpha\nu}. 
\end{equation}

Here the $K^{\ast}K$ transition form factor is given as%
\begin{equation}
F_{K^{\ast}K}^{lo}(k^{2})=\sum_{V=\rho,\omega.\phi}\frac{G_{T}C_{VK^{\ast
+}K^{-}}}{\sqrt{2}}\frac{F_{V}\ C_{V}}{3M_{K^{\ast}}}\frac{1}{k^{2}-M_{V}^{2}%
}=\frac{F_{V}\ G}{6}\left( \frac{M_{\omega}}{k^{2}-M_{\omega}^{2}}\pm \frac{%
3M_{\rho}}{k^{2}-M_{\rho}^{2}}-\frac{2M_{\phi}}{k^{2}-M_{\phi}^{2}}\right)
\end{equation}
where the upper (lower) sign corresponds to the charged (neutral) case. The
explicitly gauge invariant tensor $T_{\mu\alpha\nu}$ is given by 
\begin{equation}
T_{\mu\alpha\nu}=i\int\frac{d^{4}l}{(2\pi)^{4}}\frac{k_{\alpha}\Delta
_{\quad\mu}^{\alpha\quad\gamma\delta}(k)\epsilon_{\gamma\delta\phi\eta}%
\Delta^{\phi\eta\sigma\tau}(l)\epsilon_{\sigma\tau\alpha\nu}}{\square
_{K}\left( l+Q\right) \square_{K^{\ast}}(l)\square_{K}\left( l-k\right) }.
\label{looptensor}
\end{equation}
The calculation of this tensor, the separation of the effects at the
different scales involved in our reaction and the fixing of the necessary
substraction constants is rather involved and for the sake of clarity we
deferred it to the Appendix. We calculate this tensor in the Appendix as 
\begin{equation}
T_{\mu\alpha\nu}=\frac{16}{M_{K^{\ast}}^{2}}\frac{1}{16\pi^{2}}\left[ \left(
2+I_{G}-I_{2}+\frac{1}{2}\log\frac{m_{K}^{2}}{\mu^{2}}+\frac{Q\cdot k}{%
m_{K}^{2}}\ J_{V}\right) g_{\mu\nu}k_{\alpha}-\frac{1}{m_{K}^{2}}\ J_{V}\
(k^{2}g_{\mu\nu}-k_{\mu}k_{\nu})Q_{\alpha}\right] ,
\end{equation}
with%
\begin{eqnarray}
J_{V} & =&\int_{0}^{1}dx\int_{0}^{x}dy\frac{y(1-x)}{1-\frac{Q^{2}}{m_{K}^{2}}%
x(1-x)-\frac{2Q\cdot k}{m_{K}^{2}}(1-x)y-\frac{k^{2}}{m_{K}^{2}}y(1-y)-\frac{%
\left( m_{V}^{2}-m_{K}^{2}\right) }{m_{K}^{2}}(y-x)-i\varepsilon}
\label{JV} \\
I_{2} & =&\int_{0}^{1}dx\int_{0}^{x}dy\log[1-\frac{Q^{2}}{m_{K}^{2}}x(1-x)-%
\frac{2Q\cdot k}{m_{K}^{2}}(1-x)y-\frac{k^{2}}{m_{K}^{2}}y(1-y)-\frac{\left(
m_{V}^{2}-m_{K}^{2}\right) }{m_{K}^{2}}(y-x)-i\varepsilon]  \label{I2} \\
I_{G} & =&\int_{0}^{1}dx\log\left( 1-\frac{m_{\pi\pi}^{2}}{m_{K}^{2}}%
x(1-x)-i\varepsilon\right) \label{IG}
\end{eqnarray}
Altogether we obtain the amplitude as%
\begin{equation}
-i\mathcal{M}_{+}=\frac{-2e^{2}}{16\pi^{2}m_{K}^{2}}F_{K^{%
\ast+}K^{-}}^{lo}(k^{2})\frac{G_{T}}{\sqrt{2}M_{K^{\ast}}}\frac{t_{K\pi}^{0}%
}{\sqrt{3}}\frac{L^{\mu}}{k^{2}}\left[ I_{V}\ k_{\alpha}g_{\mu\nu}-J_{V}\
(k^{2}g_{\mu\nu}-k_{\mu}k_{\nu})Q_{\alpha}\right] \eta^{\alpha\nu}
\end{equation}
with%
\begin{equation}
I_{V}\equiv m_{K}^{2}\left( I_{G}-I_{2}+2+\frac{1}{2}\log\frac{m_{K}^{2}}{%
\mu^{2}}\right) +Q\cdot k\ J_{V}.
\end{equation}
Calculations for the amplitude $\mathcal{M}_{0}$ corresponding to neutral $%
K^{\ast}$ in the loops are are quite similar and can be obtained from $%
\mathcal{M}_{+}$ just replacing the charged transition form factor by the
neutral one due to $t_{K^{+}\pi^{+}}^{0}=t_{K^{0}\pi^{+}}^{0}$ $\equiv \frac{%
t_{K\pi}^{0}}{\sqrt{3}}$. Adding up these amplitudes we get 
\begin{equation}
-i\mathcal{M}_{V}=\frac{-2e^{2}}{16\pi^{2}m_{K}^{2}}\widetilde{F}%
_{K^{\ast}K}^{0}(k^{2})\frac{G_{T}}{\sqrt{2}M_{K^{\ast}}}\frac{t_{K\pi}^{0}}{%
\sqrt{3}}\frac{L^{\mu}}{k^{2}}\left[ I_{V}\ k_{\alpha}g_{\mu\nu}-J_{V}\
(k^{2}g_{\mu\nu}-k_{\mu}k_{\nu})Q_{\alpha}\right] \eta^{\alpha\nu}.
\end{equation}
where the isoscalar transition form factor to lowest order is given by 
\begin{equation}
\widetilde{F}_{K^{\ast}K}^{0}(k^{2})=F_{K^{\ast+}K^{-}}^{lo}(k^{2})+F_{K^{%
\ast+}K^{-}}^{lo}(k^{2})=\frac{F_{V}\ G}{3}\left( \frac{M_{\omega}}{%
k^{2}-M_{\omega}^{2}}-\frac{2M_{\phi}}{k^{2}-M_{\phi}^{2}}\right) .
\label{Fkskvmd}
\end{equation}
This amplitude can be written in terms of the conventional polarization
vector for an on-shell $\phi$ using Eqs. (\ref{keta},\ref{L12}) and $%
G_{T}=M_{\phi }M_{K^{\ast}}G$. We also replace the lowest order terms in Eq.
(\ref{Fkskvmd}) by the full transition form factor to obtain%
\begin{equation}
-i\mathcal{M}_{V}=\frac{-ie^{2}}{2\pi^{2}m_{K}^{2}}\frac{t_{K\pi}^{0}}{\sqrt{%
3}}\frac{L^{\mu}}{k^{2}}\left[ A_{V}\ L_{\mu\nu}^{(1)}+B_{V}L_{\mu\nu }^{(2)}%
\right] \eta^{\nu}
\end{equation}
with 
\begin{equation}
A_{V}=\frac{G}{4\sqrt{2}}F_{K^{\ast}K}^{0}(k^{2})I_{V},\qquad B_{V}=-\frac{%
GM_{\phi}^{2}}{4\sqrt{2}}F_{K^{\ast}K}^{0}(k^{2})J_{V}.  \label{AVBV}
\end{equation}

This contribution is proportional to the isoscalar transition form factor $%
F_{K^{\ast}K}^{0}(k^{2})$ and, similarly to the kaon form factor in the case
of kaon loops, we need a proper description of this form factor at the
energy of the reaction, which could be achieved either by a proper
unitarization of this form factor or using experimental data if they exist.
At the energy region of interest the unitarization of this form factor would
reproduce the poles of known vector resonances coupled to the $K^{\ast}K$
system. The lowest order result in Eq. (\ref{Fkskvmd}) already contains the
poles corresponding to the lowest lying vectors. The PDG list the $%
\omega(1650),$ $\phi(1680)$ and $\rho(1700)$ resonances in this energy
region, which we will call $\omega^{\prime},\phi^{\prime},\rho^{\prime}$ in
the following. In this concern it is remarkable that studies of $%
e^{+}e^{-}\rightarrow K^{0}K^{\pm}\pi^{\mp}$ at $\sqrt{s}=$ $1400-2180\
MeV$ show that this reaction is dominated by intermediate neutral $%
K^{\ast0}K^{0}$ production (with a small contribution of the charged channel
and negligible light vector meson contributions) in turn coming from
intermediate $\phi^{\prime}$ and $\rho^{\prime}$\cite{BiselloCS}. There is
no evidence for $\omega^{\prime}$ contributions in these reactions.
Furthermore, a direct measurements of the kaon form factors in $%
e^{+}e^{-}\longrightarrow K^{+}K^{-},K^{0}\overline {K}^{0}$ \cite{BiselloFF}
at $\sqrt{s}=$ $1400-2200\ MeV$ shows also evidence for contributions of $%
\phi^{\prime}$ and $\rho^{\prime}$ to the kaon form factors (again no signal
for $\omega^{\prime}$ is found here) around $1700$ $MeV$ and there is no
signal for contributions of higher vector resonances in the charged case.
Although the inclusion of such effects improves the description of the kaon
form factor around $1700\ MeV$ the values around $2.2\ GeV$ are roughly the
same as those of the unitarized charged kaon form factor \cite{OOP}. Coming
back to the $K^{\ast}K$ transition form factor, in Ref. \cite{BiselloCS} the
product 
\begin{equation}
\Gamma(\phi^{\prime}\rightarrow e^{+}e^{-})BR(\phi^{\prime}\rightarrow
K^{\ast}K)=0.39\pm0.11\ KeV,
\end{equation}
is measured, and assuming that $K^{\ast}K$ is the dominant channel for the $%
\phi^{\prime}$ meson, it allows us to extract the $\phi^{\prime}\gamma$
coupling which we write as $g_{\phi^{\prime}\gamma}=\frac{%
em_{\phi^{\prime}}^{2}}{f_{\phi^{\prime}}}$ from%
\begin{equation}
\Gamma(\phi^{\prime}\rightarrow e^{+}e^{-})=\frac{4\pi\alpha^{2}m_{\phi^{%
\prime}}}{3f_{\phi^{\prime}}^{2}}=0.39\pm0.11\ KeV,
\end{equation}
which yields $f_{\phi^{\prime}}=31$. Similarly the $\phi^{\prime}K^{\ast}K$
coupling can be extracted from the total width 
\begin{equation}
\Gamma(V\rightarrow V^{\prime}P)=\frac{g_{VV^{\prime}P}^{2}}{4\pi}|\mathbf{p|%
}^{3}
\end{equation}
which for the case at hand ($|\mathbf{p|}=462\ MeV,$ $\Gamma=150\ MeV$) and
assuming same coupling of the $\phi^{\prime}$ to $K^{\ast+}K^{-}$ and $%
K^{\ast0}K^{0}$ yields $g_{\phi^{\prime}K^{\ast}K}=2g_{\phi^{\prime}K^{\ast
0}K^{0}}=2g_{\phi^{\prime}K^{\ast+}K^{-}}=4.37\times10^{-3}\ MeV^{-1}.$

Taking into account the $\phi^{\prime}$ and $\rho^{\prime}$ contribution
introduces a factor%
\begin{equation}
\frac{g_{\phi^{\prime}K^{\ast}K}}{2f_{\phi^{\prime}}}\left( \pm\frac{3}{2}%
\frac{m_{\rho^{\prime}}^{2}}{k^{2}-m_{\rho^{\prime}}^{2}+im_{\rho^{\prime}}%
\Gamma_{\rho^{\prime}}}-\frac{m_{\phi^{\prime}}^{2}}{k^{2}-m_{\phi^{%
\prime}}^{2}+im_{\phi^{\prime}}\Gamma_{\phi^{\prime}}}\right)
\end{equation}
in the transition form factor of charged $(+)$ and neutral $(-)$ $K^{\ast}K$
in the loops. Contributions from $\rho^{\prime}$ cancel in the sum, thus the
isoscalar transition form factor is given by%
\begin{equation}
F_{K^{\ast}K}^{0}(k^{2})=\frac{F_{V}\ G}{3}\left( \frac{M_{\omega}}{%
k^{2}-M_{\omega}^{2}+im_{\omega}\Gamma_{\omega}}-\frac{2M_{\phi}}{%
k^{2}-M_{\phi}^{2}+im_{\phi}\Gamma_{\phi}}\right) -\frac{g_{\phi^{\prime
}K^{\ast}K}}{f_{\phi^{\prime}}}\left( \frac{m_{\phi^{\prime}}^{2}}{%
k^{2}-m_{\phi^{\prime}}^{2}+im_{\phi^{\prime}}\Gamma_{\phi^{\prime}}}\right). 
 \label{FKsK}
\end{equation}

Finally, taking into account both pseudoscalar and vectors in the loops we
obtain the total amplitude as%
\begin{equation}
-i\mathcal{M}=\frac{-ie^{2}}{2\pi ^{2}m_{K}^{2}}\frac{t_{K\pi }^{0}}{\sqrt{3}%
}\frac{1}{k^{2}}\overline{v}(p^{+})\gamma ^{\mu }u(p^{-})\left[ A\ L_{\mu
\nu }^{(1)}+BL_{\mu \nu }^{(2)}\right] \eta ^{\nu }
\end{equation}%
where%
\begin{eqnarray}
A& =&A_{P}+A_{V} \\
B& =&B_{P}+B_{V}
\end{eqnarray}%
with the specific functions in Eqs.(\ref{AP},\ref{BP},\ref{AVBV}). Recall these
results are valid for ingoing particles. For the numerical computations in
the following section we reverse the momenta of the final particles and
obtain%
\begin{equation}
-i\mathcal{M}=\frac{ie^{2}}{2\pi ^{2}m_{K}^{2}}\ \frac{t_{K\pi }^{0}}{\sqrt{3%
}}\frac{1}{k^{2}}\overline{v}(p^{+})\gamma ^{\mu }u(p^{-})\left[ I\ L_{\mu
\nu }^{(1)}-J\ L_{\mu \nu }^{(2)}\right] \eta ^{\nu }  \label{finalamp}
\end{equation}%
with%
\begin{eqnarray}
I& =&\frac{\sqrt{2}M_{\phi }}{2f^{2}}\left( G_{V}F_{K}^{0}(k^{2})I_{P}-(G_{V}-%
\frac{F_{V}}{2})\frac{m_{K}^{2}}{4M_{\phi }^{2}}g_{K}(m_{\pi \pi
}^{2})\right)  \label{Idecay}\\
&& -\frac{G}{4\sqrt{2}}F_{K^{\ast }K}^{0}(k^{2})\left[ Q\cdot k\
J_{V}-m_{K}^{2}\left( I_{G}-I_{2}+2+\frac{1}{2}\log \frac{m_{K}^{2}}{\mu ^{2}%
}\right) \right] \notag \\
J& =&\frac{\sqrt{2}M_{\phi }G_{V}}{2f^{2}}F_{K}^{0}(k^{2})\left( J_{P}+\frac{%
m_{K}^{2}}{4k^{2}}g_{K}\right) -\frac{GM_{\phi }^{2}}{4\sqrt{2}}F_{K^{\ast
}K}^{0}(k^{2})J_{V}.  \label{Jdecay}
\end{eqnarray}%
and for the integrals $I_{P},$ $J_{P},$ $J_{V},$ and $I_{2}$ we must use
Eqs. (\ref{IAB},\ref{JAB},\ref{JV},\ref{I2},\ref{IG}) just changing the sign of $Q\cdot k$. Also,
since our analysis include an energy region relatively far from the $\phi
^{\prime }$ peak we use in Eq. (\ref{FKsK}) an $s$-dependent width given by%
\begin{equation}
\Gamma _{\phi ^{\prime }}(s)=\frac{g_{\phi ^{\prime }K^{\ast }K}^{2}}{4\pi }%
\left( \frac{\lambda ^{\frac{1}{2}}(s,m_{K^{\ast }}^{2},m_{K}^{2})}{2\sqrt{s}%
}\right) ^{3}.
\end{equation}
with%
\begin{equation}
\lambda(m_{1}^{2},m_{2}^{2},m_{3}^{2})=(m_{1}^{2}-(m_{2}-m_{3})^{2})(m_{1}^{2}-(m_{2}+m_{3})^{2}).
\end{equation}

\section{Numerical results}

The differential cross section for this process is given as 
\begin{equation}
\frac{d\sigma}{dm_{\pi\pi}d\Omega_{Q}}=\frac{1}{\left( 2\pi\right) ^{4}}%
\frac{1}{8s^{\frac{3}{2}}}|\mathbf{Q}||\widetilde{\mathbf{p}}||\overline {%
\mathcal{M}}|^{2}.  \label{DCS}
\end{equation}
Here $\mathbf{Q}$ stands for the tri-momentum of the $\phi$ in the center of
momentum system of the reaction and $\widetilde{\mathbf{p}}$ denotes the
momentum of the final charged pion in the dipion center of momentum system%
\begin{equation}
|\mathbf{Q}|=\frac{\lambda^{\frac{1}{2}}(s,M_{\phi}^{2},m_{\pi\pi}^{2})}{2%
\sqrt{s}},\qquad|\widetilde{\mathbf{p}}|=\frac{\lambda^{\frac{1}{2}%
}(m_{\pi\pi}^{2},m_{\pi}^{2},m_{\pi}^{2})}{2m_{\pi\pi}},
\end{equation}
where we neglect terms proportional to $m_{e}^{2}$ . A straightforward
calculation yields%
\begin{eqnarray}
|\overline{\mathcal{M}}|^{2} & =&\frac{1}{4}\sum_{pol}|\mathcal{M}%
|^{2}=|C|^{2}\,\left[ |I|^{2}\frac{1\,}{2}\,\left( M_{\phi}^{2}+|\mathbf{Q}%
|^{2}\,x^{2}+{\omega}^{2}\right) -\,2 Re(IJ^{\ast }\,)\,\sqrt{s}%
\,\omega\,+|J|^{2}\frac{s}{2\,M_{\phi}^{2}}\,\left( M_{\phi }^{2}-|\mathbf{Q}%
|^{2}\,x^{2}+{\omega}^{2}\right) \right] \\
& =&|C|^{2}\,\left[ |I|^{\,2}\frac{1\,}{2}\left( M_{\phi}^{2}(1-x^{2})+{\omega%
}^{2}\,(1+x^{2})\right) -\,2 Re(IJ^{\ast}\,)\sqrt {s}\,\omega\,+|J|^{2}%
\frac{s}{2\,M_{\phi}^{2}}\,\left( M_{\phi}^{2}(1+x^{2})+{\omega}%
^{2}\,(1-x^{2})\right) \right]
\end{eqnarray}
where $x=\cos\theta$ with $\theta$ the $\phi$-beam angle, ${\omega}$ the $%
\phi$ energy 
\begin{equation}
\omega=\frac{s+M_{\phi}^{2}-m_{\pi\pi}^{2}}{2\sqrt{s}}
\end{equation}
and $C$ stands for the global factor 
\begin{equation}
C=\frac{ie^{2}}{2\pi^{2}m_{K}^{2}}\ \frac{t_{K\pi}^{0}}{\sqrt{3}}.
\end{equation}
Integrating the solid angle we get%
\begin{equation}
\int|\overline{\mathcal{M}}|^{2}d\Omega_{Q}=\frac{4\pi}{3}|C|^{2}\,\left[
|I|^{2}\,\left( M_{\phi}^{2}+2{\omega}^{2}\,\,\right) -\,6 Re%
(IJ^{\ast}\,)\sqrt{s}\,\omega\,+|J|^{2}\frac{s}{\,M_{\phi}^{2}}\,\left(
2M_{\phi}^{2}+{\omega}^{2}\,\right) \right]
\end{equation}
The dipion spectrum is finally given as

\begin{equation}
\frac{d\sigma}{dm_{\pi\pi}}=\frac{\alpha^{2}}{24\pi^{5}m_{K}^{4}}\frac{|%
\mathbf{Q}||\widetilde{\mathbf{p}}|}{s^{\frac{3}{2}}}\ \frac{|t_{K\pi
}^{0}|^{2}}{3}h(s,m_{\pi\pi})  \label{spectrum}
\end{equation}
where%
\begin{equation}
h(s,m_{\pi\pi})=|I|^{2}\,\left( M_{\phi}^{2}+2{\omega}^{2}\,\,\right) -\,6%
 Re(IJ^{\ast}\,)\sqrt{s}\,\omega\,+|J|^{2}\frac{s}{\,M_{\phi}^{2}}%
\,\left( 2M_{\phi}^{2}+{\omega}^{2}\right) .
\end{equation}
We evaluate numerically the integrals and the differential cross section. We 
are interested in dipion energies $m_{\pi\pi}$ close to the 
$f_{0}(980)$ mass in whose case the $K\bar{K}\to\pi\pi$ scattering between the 
kaons in the loops and the final pions takes place at this energy independently 
of the value of $\sqrt{s}$ and of the momenta in the loops. As a consequence,  
when replacing the lowest order terms for this amplitude by the unitarized amplitude 
as proposed in Section II, we can safely use the results 
of \cite{OO,OOPel} and take a renormalization scale $\mu=1.2$ $GeV$ for the 
function $G_{K}$ \cite{OOPel}, in spite of the fact that the reaction takes place 
at a much higher energy $\sqrt{s}\geq 2 \, GeV$. The unitarized amplitudes 
naturally contain the scalar poles and there is no need to include explicitly 
these degrees of freedom in the calculation. For the kaon form factor we use 
the unitarized version calculated in Ref. \cite{OOP}, the values obtained in 
this analytic form reproduce the direct
measurements of the charged kaon form factor at the energy region of
interest \cite{BiselloFF}. 

\begin{center}
\begin{figure}[ptb]
\begin{center}
\includegraphics[
natheight=8.25in,
natwidth=9.48in,
height=4.50in,
width=5.25in
]{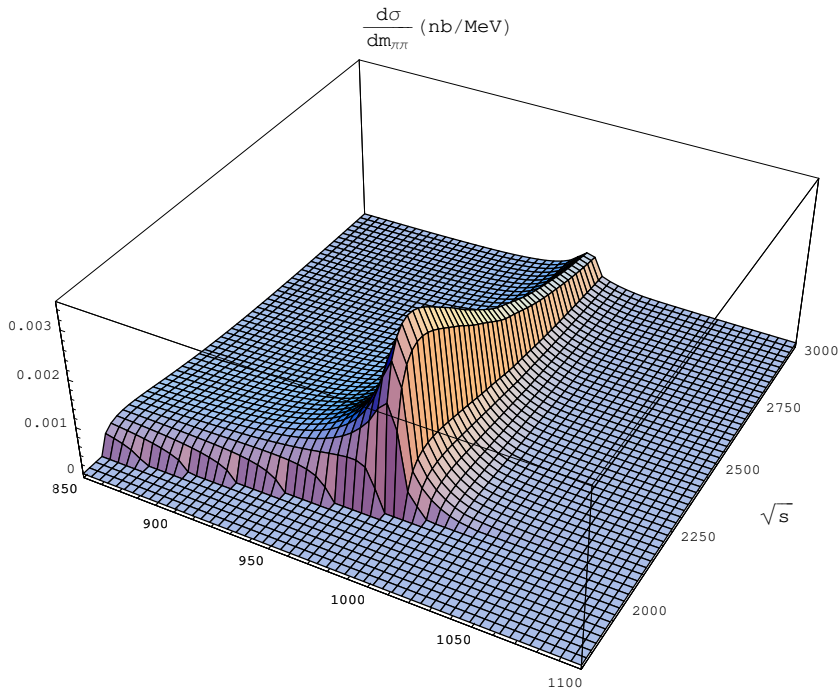}
\end{center}
\caption{Differential cross section as a function of the dipion invariant
mass and of the center of mass energy.}
\label{dsig}
\end{figure}
\end{center}

Using the physical masses and coupling constants $%
m_{K}=495,$ $m_{\phi}=1019.4$, $\alpha=1/137$, $G_{V}=53MeV$, $F_{V}=154MeV$%
, $f_{\pi}=93~MeV$, and $G=0.016MeV^{-1}$ in Eq. (\ref{spectrum}) we obtain
the spectrum shown in Figs. (\ref{dsig}) where the presence of the $%
f_{0}(980)$ is well visible. This is a consequence of the fact that the $%
f_{0}(980)$ poles are well reproduced in the unitarization of meson meson $s$%
-wave isoscalar amplitudes present in our calculation. The $\sqrt{s}$
dependence in the differential cross section is dominated by the phase space
factor in the lower energy region (the opening of the $\phi f_{0}$ channel)
and the lowering beyond the $\phi f_{0}$ threshold is dictated by the form
factors.

\begin{center}
\begin{figure}[ptb]
\begin{center}
\includegraphics[
natheight=10.0in,
natwidth=10.0in,
height=5.15in,
width=5.18in
]{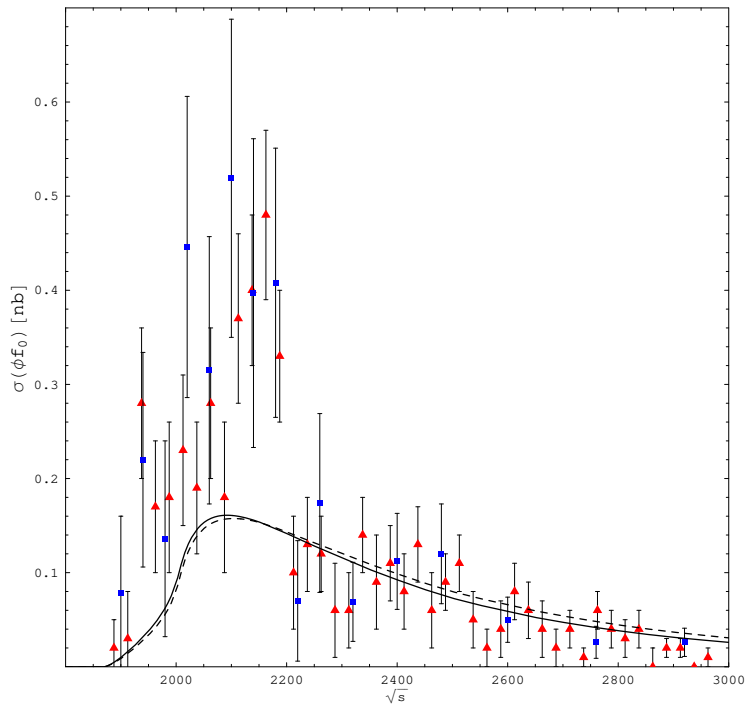}
\end{center}
\caption{Cross section for $e^{+}e^{-}\rightarrow\protect\phi\left[ \protect%
\pi\protect\pi\right] _{I=0}$, integrated in the $m_{\protect\pi\protect\pi%
}=850-1100~ MeV$ range, as a function of $\protect\sqrt{s}$ including all contributions.
Experimental points from Ref. \protect\cite{BBX}, triangled (boxed) points
correspond to charged (neutral) pions.}
\label{sig}
\end{figure}
\end{center}

Next we integrate $m_{\pi \pi }$ from $850MeV$ to $1100~MeV$ following the
cuts implemented in \cite{BBX} . The obtained cross section is shown in 
Fig (\ref{sig}) (solid curve) where we also show the experimental points quoted
in Ref. \cite{BBX}. We must remark that all the parameters in Eqs. (\ref{Idecay},
\ref{Jdecay}) have been fixed in advance and in this sense there are no free
parameters in our calculations. We should note that in the loops with
pseudoscalars there is a term that has no form factor. At low photon
virtualities this term is small and its extrapolation to high $k^{2}$
requires to dress it with a form factor which does not come from the
Lagrangians that we are using. Thus some uncertainty should be accepted at
this point. However, we find numerically that the contributions of the loops
with pseudoscalars is far smaller than the contributions of the vector meson
loops (by themselves one order of magnitude smaller close to the $\phi f_{0}$
threshold) but through interference with vector meson loops they become more
relevant). The effect of the term with no form factor is shown in Fig (\ref{sig}) 
where we plotted the cross section as a function of $\sqrt{s}$ in the
case when this term is absent (solid line) and dressed with the kaon form
factor (dashed line). As we can see, the effect of this term is negligible when 
dressed with the kaon form factor.

The elaborate theoretical study carried out in this paper, using standard
tools to produce the $\phi f_{0}(980)$ has succeeded in reproducing the bulk
of the experimental data as a function of the energy. Yet, the theory,
producing reasonable numbers around $\sqrt{s}=2000$ $MeV$ and beyond $2300$ $%
MeV$, fails to provide the right strenght in the region around $2150$ $MeV$
where a peak appears in the data. There is no way, within our theoretical
framework, with reasonable changes of the parameters within existing
uncertainties, to obtain this peak. As a consequence of it, we are inclined
to conclude, following the lines of Ref. \cite{BBX}, that there is a $1^{--%
\text{ }}$meson resonance around $2150$ $MeV$ coupling strongly to $\phi
f_{0}(980)$, as also concluded in \cite{BBX}. In as much as our theoretical
results provides a "background" very similar to the one assumed there, our
conclusions about the resonance are the same as in \cite{BBX} and we refrain
from repeating the same analysis leading to the properties of the new
resonance. Recalling the result from \cite{BBX}, the resonance has a mass of 
$M_{R}=2175$ $MeV$, a width of $\Gamma=58$ $MeV$ and quantum numbers $1^{--%
\text{ }}$as the photon.

From the theoretical point of view such a resonance is a real challenge
since their properties are not predicted by ordinary quark models hinting to
a possible exotic character \cite{structure}. 

\section{Summary and conclusions}

We studied electron-positron annihilation into $\phi \pi \pi $ for pions in
an isoscalar s-wave. We find the tree level contributions induced via $%
\omega -\phi $ mixing negligible. At one loop level, using the vector mesons
interactions arising in $R\chi PT$ we show the cancellation of the
contributions coming from the off-shell parts of the meson-meson amplitudes
in the calculation of the kaon loops. The on-shell parts are iterated to
obtain the unitarized meson-meson amplitudes. We obtain contributions
proportional to these amplitudes and to the lowest order terms of the kaon
form factors. In addition, we find a term with the unitarized meson-meson
amplitudes but without the kaon form factors. The effect of the latter is
negligible when dressed with the kaon form factor. The photon exchanged in 
$e^{+}e^{-}\rightarrow \phi \pi \pi $ is highly virtual and the proper
description of this process requires to use the full kaon form factors.
Thus, instead of the lowest order terms arising in the calculation we use
the full form factor as calculated in $U\chi PT$ \cite{OOP}.

The high virtuality of the exchanged photon makes the excitation of higher
mass states likely. We calculate the excitation of $K^{\ast}K$ states with
rescattering of kaons into the final pions. This contribution is calculated
using $U\chi PT$ supplemented with the anomalous term describing $VVP$
interactions. There are two different energy scales involved in the
reaction: $M_{\phi},m_{\pi\pi}\approx\Lambda$ and $\sqrt{k^{2}}\gtrsim2GeV$
and we perform a clear separation of the effects at these scales. It is
shown that the only substraction constant required is the one associated to
the meson-meson scattering. The formalism naturally yields the contribution
from light vector mesons to the $K^{\ast}K$ transition form factor. However,
the proper description of this form factor at the energy of the reaction
requires to include contributions from heavy mesons, which are extracted
from the data on $e^{+}e^{-}\longrightarrow K^{0}K^{\pm}\pi^{\mp}$ at $\sqrt{%
s}=$ $1400-2180\ MeV$ \cite{BiselloCS}. All the parameters entering the
calculation have been fixed in advance and there is no freedom in their
choice. For the differential cross section we find a peak in $m_{\pi\pi}$
around the $f_{0}(980)$ as in the experiment \cite{BBX}. We select the $\phi$
$f_{0}(980)$ events imposing the cuts used in the analysis of Ref. \cite{BBX}%
. The corresponding cross section as a function of the $e^{+}e^{-}$ energy
describes satisfactorily the experimental data except for a narrow peak
around $2150$ $MeV$, yielding support to the existence of a $1^{--}$
resonance above the $\phi f_{0}(980)$ threshold whose structure started to be 
debated and seems to be non-conventional \cite{structure}. 
On the other hand, the
description of the peaks of $m_{\pi\pi}$ around the $f_{0}(980)$ resonance,
as well as the agreement with data on total cross sections (up to the signal
of the new resonance), without the explicit introduction of the $f_{0}(980)$
state, provides extra support for the $f_{0}(980)$ as being dynamically
generated from the interaction of pseudoscalar mesons in coupled channels. 

\begin{acknowledgments}
We wish to thank M. J. Vicente-Vacas for usefull discussions and a critical
reading of the manuscript. This work is partly supported by DGICYT contract
number FIS2006-03438 and the Generalitat Valenciana. This research is part
of the EU Integrated Infrastructure Initiative Hadron Physics Project under
contract number RII3-CT-2004-506078. The work of M. Napsuciale was also
supported by CONACyT-M\'{e}xico under project CONACyT-50471-F. C. A.
Vaquera-Araujo wish to acknowledge support by CONACyT-M\'{e}xico under the
Mixed Grants Program.
\end{acknowledgments}

\section{Appendix}

We use dimensional regularization to calculate the loop tensor in Eq. (\ref%
{looptensor}) which in dimension $d=4-2\varepsilon$ reads 
\begin{equation}
T_{\mu\alpha\nu}=i\mu^{2\varepsilon}\int\frac{d^{d}l}{(2\pi)^{d}}\frac {%
g_{\mu\beta}k_{\alpha}\Delta^{\alpha\beta\gamma\delta}(k)\epsilon
_{\gamma\delta\phi\eta}\Delta^{\phi\eta\sigma\tau}(l)\epsilon_{\sigma
\tau\alpha\nu}}{\square_{K}\left( l+Q\right)
\square_{K^{\ast}}(l)\square_{K}\left( l-k\right) },
\end{equation}
where $\mu$ stands for the renormalization scale. Using Eq.(\ref{simple})
and the anti-symmetry of the Levi-Civita tensors we get%
\begin{eqnarray}
g_{\mu\beta}k_{\alpha}\Delta^{\alpha\beta\gamma\delta}(k)\epsilon
_{\gamma\delta\phi\eta} & =&\left( g_{\mu}^{\gamma}k^{\delta}-g_{\mu
}^{\delta}k^{\gamma}\right) \epsilon_{\gamma\delta\phi\eta}=2k^{\delta
}\epsilon_{\mu\delta\phi\eta} \\
\epsilon_{\mu\delta\phi\eta}\Delta^{\phi\eta\sigma\tau}(l)\epsilon_{\sigma
\tau\alpha\nu} & =&\frac{2}{m_{K^{\ast}}^{2}}\left( \left(
l^{2}-m_{K^{\ast}}^{2}\right)
\epsilon_{\mu\delta\phi\eta}\epsilon_{\quad\alpha
\nu}^{\phi\eta}+2\epsilon_{\mu\delta\phi\eta}\epsilon_{\sigma\quad\alpha\nu
}^{\quad\phi}l^{\eta}l^{\sigma}\right)
\end{eqnarray}
which allows us to split the loop tensor as%
\begin{equation}
T_{\mu\alpha\nu}=\frac{4}{m_{K^{\ast}}^{2}}\left(
T_{\mu\alpha\nu}^{(1)}+T_{\mu\alpha\nu}^{(2)}\right) ,  \label{tdef}
\end{equation}
where%
\begin{eqnarray}
T_{\mu a\nu}^{(1)} & =&i\mu^{2\varepsilon}\int\frac{d^{d}l}{(2\pi)^{d}}\frac{%
k^{\delta}\epsilon_{\mu\delta\phi\eta}\epsilon_{\quad\alpha\nu}^{\phi\eta}}{%
\square\left( l+Q\right) \square\left( l-k\right) }=(d-3)(d-2)\left(
k_{\alpha}g_{\mu\nu}-k_{\nu}g_{\mu\alpha}\right) \mu^{2\varepsilon}\int\frac{%
d^{d}l}{(2\pi)^{d}}\frac{i}{\square\left( l+Q\right) \square\left(
l-k\right) } \\
T_{\mu\alpha\nu}^{(2)} & =&2i\mu^{2\varepsilon}\int\frac{d^{d}l}{(2\pi)^{d}}%
\frac{k^{\delta}\epsilon_{\mu\delta\phi\eta}\epsilon_{\sigma\quad\alpha\nu
}^{\quad\phi}l^{\eta}l^{\sigma}}{\square_{K}\left( l+Q\right) \square
_{K^{\ast}}(l)\square_{K}\left( l-k\right) }.
\end{eqnarray}
A straightforward calculation yields%
\begin{equation}
T_{\mu a\nu}^{(1)}=4k_{\alpha}g_{\mu\nu}\left[ \frac{3}{16\pi^{2}}%
+G_{K}(m_{\pi\pi}^{2})\right] 
\end{equation}
where we used $\eta^{\alpha\nu}=-\eta^{\nu\alpha}$. Notice that we get a
constant contribution coming from the contraction of the Levi-Civita tensors
in dimension $d$ besides the conventional loop function $G_{K}$.

The second loop tensor contains two different scales: $M_{\phi},m_{\pi\pi
}\approx\Lambda$ and $\sqrt{k^{2}}\gtrsim2GeV$ and we must ensure a clean
separation of the effects at these scales and the correct estimate of the
corresponding substraction constants. With this aim we perform a
decomposition of this tensor in terms of scalar integrals. The tensor
integral%
\begin{equation}
C^{\eta\sigma}=i\mu^{2\varepsilon}\int\frac{d^{d}l}{(2\pi)^{d}}\frac{l^{\eta
}l^{\sigma}}{\square_{K}\left( l+Q\right) \square_{K^{\ast}}(l)\square
_{K}\left( l-k\right) }
\end{equation}
can be decomposed as%
\begin{equation}
C_{\eta\sigma}=C_{00}g_{\eta\sigma}+C_{11}Q_{\eta}Q_{\sigma}+C_{12}(Q_{\eta
}k_{\sigma}+k_{\eta}Q_{\sigma})+C_{22}k_{\eta}k_{\sigma}.
\end{equation}
We will be interested only in the coefficients of $g_{\eta\sigma}$ and $%
Q_{\eta}k_{\sigma}$ since the remaining terms give vanishing contributions
to the process at hand. Contracting with $g^{\eta\sigma},Q^{\eta}$ and $%
k^{\eta}$ we get the following equations for the coefficients%
\begin{eqnarray}
d\text{ }C_{00}+Q^{2}C_{11}+2Q\cdot kC_{12}+k^{2}C_{22} & =&G_{K}(m_{\pi\pi
}^{2})+M_{V}^{2}C_{0}\equiv R_{00} \\
C_{00}+Q^{2}C_{11}+Q\cdot kC_{12} & =&\frac{1}{2}\left( \frac{1}{2}%
G_{K}(m_{\pi\pi}^{2})-\left( Q^{2}+\Delta^{2}\right) C_{1}\right) \equiv
R_{11} \\
Q^{2}C_{12}+Q\cdot kC_{22} & =&\frac{1}{2}\left( V_{1}(k^{2})-\frac{1}{2}%
G_{K}(m_{\pi\pi}^{2})-\left( Q^{2}+\Delta^{2}\right) C_{2}\right) \equiv
R_{12} \\
C_{00}+Q\cdot kC_{12}+k^{2}C_{22} & =&\frac{1}{2}\left( \left(
k^{2}+\Delta^{2}\right) C_{2}+\frac{1}{2}G_{K}(m_{\pi\pi}^{2})\right) \equiv
R_{22},
\end{eqnarray}
where $\Delta^{2}=M_{V}^{2}-m_{K}^{2},$ $C_{0}$ stands for the finite scalar
integral%
\begin{equation}
C_{0}=\mu^{2\varepsilon}\int\frac{d^{d}l}{(2\pi)^{d}}\frac{i}{\square
_{K}\left( l+Q\right) \square_{K^{\ast}}(l)\square_{K}\left( l-k\right) },
\end{equation}
and $C_{1},C_{2}$ stand for the coefficients of the decomposition of the
vector integral 
\begin{equation}
C_{\sigma}=i\mu^{2\varepsilon}\int\frac{d^{d}l}{(2\pi)^{d}}\frac{l_{\sigma}}{%
\square_{K}\left( l+Q\right) \square_{K^{\ast}}(l)\square_{K}\left(
l-k\right) }=C_{1}Q_{\sigma}+C_{2}k_{\sigma}.
\end{equation}
It can be easily shown that $C_{1}$ and $C_{2}$ are finite. The functions $%
V_{1}$ and $V$ are given by%
\begin{eqnarray}
V_{1}(k^{2}) & =&\frac{1}{2}\left[ V(k^{2})+\frac{\Delta^{2}}{k^{2}}\left(
V(k^{2})-V(0)\right) \right] \\
V(k^{2}) & =&\mu^{2\varepsilon}\int\frac{d^{d}l}{(2\pi)^{d}}\frac{i}{%
\square_{K^{\ast}}\left( l\right) \square\left( l-k\right) }.
\end{eqnarray}
The required coefficients read%
\begin{eqnarray}
C_{00} & =&\frac{1}{d-2}\left( R_{00}-R_{11}-R_{22}\right) \\
C_{12} & =&\frac{1}{d-2}\frac{1}{\left( Q\cdot k\right) ^{2}-Q^{2}k^{2}}\left[
Q\cdot k\left( -R_{00}+R_{11}+3R_{22}\right) -2k^{2}R_{12}\right] .
\end{eqnarray}
Explicitly 
\begin{eqnarray}
C_{00} & =&\frac{1}{2(d-2)}\left[ G_{K}(m_{\pi%
\pi}^{2})+2M_{V}^{2}C_{0}+Q^{2}C_{1}-k^{2}C_{2}+\Delta^{2}(C_{1}-C_{2})%
\right] , \\
C_{12} & =&-\frac{1}{4\left( \left( Q\cdot k\right) ^{2}-Q^{2}k^{2}\right) }%
\left\{ 2M_{V}^{2}Q\cdot kC_{0}+k^{2}\left(
V(k^{2})-G_{K}(m_{\pi\pi}^{2})\right) +\Delta^{2}\left( V(k^{2})-V(0)\right)
\right. \nonumber\\
& &+\left. Q\cdot k\left( Q^{2}+\Delta^{2}\right) C_{1}-\left( 3Q\cdot
k\left( k^{2}+\Delta^{2}\right) +2k^{2}\left( Q^{2}+\Delta^{2}\right)
\right) C_{2}\right\} .
\end{eqnarray}
Notice that the dependence of the integrals on the two different scales ( $%
k^{2}$ and $Q^{2},m_{\pi\pi}^{2}$ ) involved in the process have been neatly
separated. Furthermore, divergences in $V(k^{2})-V(0)$ and $%
V(k^{2})-G_{K}(m_{\pi\pi}^{2})$ cancel out rendering $C_{12}$ finite as
expected. In contrast $C_{00}$ is divergent \textit{but its divergent term
appears in }$G_{K}(m_{\pi\pi}^{2})$\textit{\ }whose finite part has already
been matched to the cutoff regularized integral. As a final result we obtain
that effects involving the scale $k^{2}$ are finite and the only
substraction constant required is the one in the loop integral associated to
the meson-meson scattering.

Contracting the Levi-Civita tensors ( in dimension $d$ ) we obtain 
\begin{eqnarray}
T_{\mu\alpha\nu}^{(2)} & =&-2\left[ Q\cdot k\ C_{12}+\left( d-2\right) C_{00}%
\right] \left( d-3\right) \left( g_{\mu\nu}k_{\alpha}-g_{\mu\alpha
}k_{\nu}\right) -2C_{12}\ \left( d-3\right) \left( k^{2}g_{\mu\alpha
}-k_{\mu}k_{\alpha}\right) Q_{\nu} \nonumber\\
&& +2C_{12}\ \left( d-3\right) \left( k^{2}g_{\mu\nu}-k_{\mu}k_{\nu }\right)
Q_{\alpha}.
\end{eqnarray}
The anti-symmetry of $\eta^{\alpha\nu}$ allows to rewrite this tensor as%
\begin{equation}
T_{\mu\alpha\nu}^{(2)}=-4\left[ Q\cdot k\ C_{12}+\left( d-2\right) C_{00}%
\right] \left( d-3\right) g_{\mu\nu}k_{\alpha}+4C_{12}\ \left( d-3\right)
\left( k^{2}g_{\mu\nu}-k_{\mu}k_{\nu}\right) Q_{\alpha}
\end{equation}
For the piece containing the divergent integral $C_{00}$ we obtain 
\begin{equation}
-4\left( d-3\right) \left( d-2\right) C_{00}=-2\left[ \frac{2}{\left(
4\pi\right) ^{2}}+G_{K}(m_{\pi%
\pi}^{2})+2M_{V}^{2}C_{0}+Q^{2}C_{1}-k^{2}C_{2}+\Delta^{2}(C_{1}-C_{2})%
\right] .
\end{equation}
The constant term in this equation comes from the dimensional factor $d-3$
which in turn arises from the contraction of the Levi-Civita tensors in
dimension $d$.

In the numerical computation it is easier to work with these integrals
written in terms of Feynman parameters. In order to calculate $%
T_{\mu\alpha\nu}^{(2)}$ we use the following Feynman parametrization%
\begin{equation}
\frac{1}{\alpha\beta\gamma}=2\int_{0}^{1}dx\int_{0}^{x}dy\frac{1}{\left[
\alpha+(\beta-\alpha)x+(\gamma-\beta)y\right] ^{3}}
\end{equation}
with 
\begin{equation}
\alpha=\left( l+Q\right)
^{2}-m_{K}^{2}+i\varepsilon,\quad\beta=l^{2}-m_{V}^{2}+i\varepsilon,\quad%
\gamma=\left( l-k\right) ^{2}-m_{K}^{2}+i\varepsilon
\end{equation}
After some algebra we get the term contributing to our process as%
\begin{equation}
T_{\mu\alpha\nu}^{(2)}=4~\epsilon_{\mu\delta\phi\eta}\epsilon_{\sigma
\quad\alpha\nu}^{\quad\phi}k^{\delta}\mu^{2\varepsilon}i\int_{0}^{1}dx\int
_{0}^{x}dy\int\frac{d^{d}r}{(2\pi)^{d}}\frac{r^{\eta}r^{\sigma}-(1-x)yQ^{%
\eta }k^{\sigma}}{\left[ r^{2}-\widetilde{m}\right] ^{3}},
\end{equation}
where%
\begin{equation}
\widetilde{m}^{2}=m_{K}^{2}-Q^{2}x(1-x)-2Q\cdot k(1-x)y-k^{2}y(1-y)-\left(
m_{V}^{2}-m_{K}^{2}\right) (y-x)-i\varepsilon,
\end{equation}
A comparison with%
\begin{equation}
T_{\mu\alpha\nu}^{(2)}=2\epsilon_{\mu\delta\phi\eta}\epsilon_{\sigma
\quad\alpha\nu}^{\quad\phi}k^{\delta}[C_{00}\ g^{\eta\sigma}+C_{12}\ Q^{\eta
}k^{\sigma}]
\end{equation}
allows us to identify 
\begin{eqnarray}
C_{00} & =&\frac{2}{d}\mu^{2\varepsilon}i\int_{0}^{1}dx\int_{0}^{x}dy\int%
\frac{d^{d}r}{(2\pi)^{d}}\frac{r^{2}}{\left[ r^{2}-\widetilde {m}\right] ^{3}%
}, \\
C_{12} & =&-2\mu^{2\varepsilon}i\int_{0}^{1}dx\int_{0}^{x}dy\int\frac{d^{d}r}{%
(2\pi)^{d}}\frac{(1-x)y}{\left[ r^{2}-\widetilde{m}^{2}\right] ^{3}}.
\end{eqnarray}
The $C_{12}$ integral is finite thus we can set $d=4$ wherever it appears to
obtain 
\begin{equation}
C_{12}=-\frac{1}{16\pi^{2}m_{K}^{2}}J_{V}  \notag
\end{equation}
with%
\begin{equation}
J_{V}\equiv\int_{0}^{1}dx\int_{0}^{x}dy\frac{(1-x)y}{1-\frac{Q^{2}}{m_{K}^{2}%
}x(1-x)-\frac{2Q\cdot k}{m_{K}^{2}}(1-x)y-\frac{k^{2}}{m_{K}^{2}}y(1-y)-%
\frac{\left( m_{V}^{2}-m_{K}^{2}\right) }{m_{K}^{2}}(y-x)-i\varepsilon}.
\end{equation}
As to the term containing the divergent integral $C_{00}$ we obtain 
\begin{equation}
-8\left( 1-\varepsilon\right) \left( 1-2\varepsilon\right) C_{00}=-\frac{2}{%
16\pi^{2}}\left( a(\mu)+3+\log\frac{m_{K}^{2}}{\mu^{2}}+2I_{2}\right)
\end{equation}
with $a(\mu)$ the substraction constant of $G(m_{\pi\pi}^{2})$ and 
\begin{equation}
I_{2}=\int_{0}^{1}dx\int_{0}^{x}dy\log[1-\frac{Q^{2}}{m_{K}^{2}}x(1-x)-\frac{%
2Q\cdot k}{m_{K}^{2}}(1-x)y-\frac{k^{2}}{m_{K}^{2}}y(1-y)-\frac{\left(
m_{V}^{2}-m_{K}^{2}\right) }{m_{K}^{2}}(y-x)-i\varepsilon]
\end{equation}
Summarizing, the tensors $T_{\mu\nu}^{(1)},T_{\mu\nu}^{(2)}$ are given by%
\begin{eqnarray}
T_{\mu\alpha\nu}^{(1)} & =&\frac{4}{16\pi^{2}}\left[ 4+\log\frac{m_{K}^{2}}{%
\mu^{2}}+I_{G}\right] k_{\alpha}g_{\mu\nu} \\
T_{\mu\alpha\nu}^{(2)} & =&\frac{4}{16\pi^{2}}\left[ \left( \frac{Q\cdot k}{%
m_{K}^{2}}J_{V}-\frac{1}{2}\left( 4+\log\frac{m_{K}^{2}}{\mu^{2}}\right)
-I_{2}\right) k_{\alpha}\ g_{\mu\nu}-\frac{1}{m_{K}^{2}}\ J_{V}\
(k^{2}g_{\mu\nu}-k_{\mu}k_{\nu})Q_{\alpha}\right] ,
\end{eqnarray}
thus from Eq.(\ref{tdef})\textit{\ }we get 
\begin{equation}
T_{\mu\alpha\nu}=\frac{16}{M_{K^{\ast}}^{2}}\frac{1}{16\pi^{2}}\left[ \left(
2+I_{G}-I_{2}+\frac{1}{2}\log\frac{m_{K}^{2}}{\mu^{2}}+\frac{Q\cdot k}{%
m_{K}^{2}}\ J_{V}\right) g_{\mu\nu}k_{\alpha}-\frac{1}{m_{K}^{2}}\ J_{V}\
(k^{2}g_{\mu\nu}-k_{\mu}k_{\nu})Q_{\alpha}\right] .
\end{equation}


\begin{thebibliography}{99}
\bibitem{daphne} %\cite{Benayoun:1999hm}
%\bibitem{Benayoun:1999hm}
M.~Benayoun, S.~I.~Eidelman, V.~N.~Ivanchenko and Z.~K.~Silagadze, 
%``Spectroscopy at B-factories using hard photon emission,''
Mod.\ Phys.\ Lett.\ A \textbf{14}, 2605 (1999) [arXiv:hep-ph/9910523]. 
%%CITATION = MPLAE,A14,2605;%%
%\cite{Denig:2006kj}
%\bibitem{Denig:2006kj}
A.~Denig, %``The radiative return: A review of experimental results,''
Nucl.\ Phys.\ Proc.\ Suppl.\ \textbf{162}, 81 (2006) [arXiv:hep-ex/0611024]; 
%%CITATION = NUPHZ,162,81;%%
%\cite{Leone:2006bm}
%\bibitem{Leone:2006bm}
D.~Leone [KLOE Collaboration], 
%``Measuring The Pion Form Factor Via Radiative Return At Large Photon Angles
%With Kloe,''
Nucl.\ Phys.\ Proc.\ Suppl.\ \textbf{162}, 95 (2006); 
%%CITATION = NUPHZ,162,95;%%
%\cite{Czyz:2005as}
%\bibitem{Czyz:2005as}
H.~Czyz, A.~Grzelinska, J.~H.~Kuhn and G.~Rodrigo, 
%``Electron positron annihilation into three pions and the radiative
%return,''
Eur.\ Phys.\ J.\ C \textbf{47}, 617 (2006) [arXiv:hep-ph/0512180]; 
%%CITATION = EPHJA,C47,617;%%
%\cite{Denig:2005eb}
%\bibitem{Denig:2005eb}
A.~G.~Denig [KLOE Collaboration], 
%``Measurement of the hadronic cross section via radiative return at DAPHNE,''
Int.\ J.\ Mod.\ Phys.\ A \textbf{20}, 1935 (2005). 
%%CITATION = IMPAE,A20,1935;%%
%\cite{Kuhn:2004zt}
%\bibitem{Kuhn:2004zt}
J.~H.~Kuhn, 
%``The radiative return at phi and B factories: A status report,''
Eur.\ Phys.\ J.\ C \textbf{33}, S659 (2004). %%CITATION = EPHJA,C33,S659;%%
%\cite{Muller:2004mb}
%\bibitem{Muller:2004mb}
S.~E.~Muller [KLOE Collaboration], 
%``Measurement of the e+ e- hadronic cross section at DAPHNE via radiative
%return,''
Nucl.\ Phys.\ Proc.\ Suppl.\ \textbf{126}, 335 (2004); 
%%CITATION = NUPHZ,126,335;%%
%\cite{Denig:2002ps}
%\bibitem{Denig:2002ps}
A.~G.~Denig \textit{et al.} [the KLOE Collaboration], 
%``Measuring the hadronic cross section via radiative return,''
Nucl.\ Phys.\ Proc.\ Suppl.\ \textbf{116}, 243 (2003)
[arXiv:hep-ex/0211024]. %%CITATION = NUPHZ,116,243;%%

\bibitem{BaBar} %\cite{Aubert:2004kj}
%\bibitem{Aubert:2004kj}
B.~Aubert \textit{et al.} [BABAR Collaboration], 
%``Study of $e^+e^- \to \pi^+ \pi^- \pi^0$ process using initial state
%radiation  with BaBar,''
Phys.\ Rev.\ D \textbf{70}, 072004 (2004) [arXiv:hep-ex/0408078]. 
%%CITATION = PHRVA,D70,072004;%%
%\cite{Aubert:2005eg}
%\bibitem{Aubert:2005eg}
B.~Aubert \textit{et al.} [BABAR Collaboration], 
%``The $e^+e^- \to \pi^+ \pi^- \pi^+ \pi^-$, $K^+ K^- \pi^+ \pi^-$, and $K^+
%K^- K^+ K^-$ cross sections at center-of-mass energies 0.5-GeV - 4.5-GeV
%measured with  initial-state radiation,''
Phys.\ Rev.\ D \textbf{71}, 052001 (2005) [arXiv:hep-ex/0502025]. 
%%CITATION = PHRVA,D71,052001;%%
%\cite{Aubert:2005cb}
%\bibitem{Aubert:2005cb}
B.~Aubert \textit{et al.} [BABAR Collaboration], 
%``A study of e+ e- --> p anti-p using initial state radiation with BABAR,''
Phys.\ Rev.\ D \textbf{73}, 012005 (2006) [arXiv:hep-ex/0512023]. 
%%CITATION = PHRVA,D73,012005;%%
%\cite{Aubert:2006jq}
%\bibitem{Aubert:2006jq}
B.~Aubert \textit{et al.} [BABAR Collaboration], 
%``The $e^+e^- \to 3(\pi^+ \pi^-), 2(\pi^+ \pi^- \pi^0)$ and $K^+ K^- 2(\pi^+
%\pi^-)$ cross  sections at center-of-mass energies from production threshold
%to  4.5-GeV measured with initial-state radiation,''
Phys.\ Rev.\ D \textbf{73}, 052003 (2006) [arXiv:hep-ex/0602006]. 
%%CITATION = PHRVA,D73,052003;%%
%\cite{BaldiniFerroli:2006ma}
%\bibitem{BaldiniFerroli:2006ma}
R.~Baldini Ferroli [BaBar Collaboration], 
%``Proton form factors and related processes in BaBar by ISR,''
Int.\ J.\ Mod.\ Phys.\ A \textbf{21}, 5565 (2006); 
%%CITATION = IMPAE,A21,5565;%%
B.~A.~Shwartz [Belle Collaboration], 
%``First steps to radiative return studies at Belle,''
Nucl.\ Phys.\ Proc.\ Suppl.\ \textbf{144}, 245 (2005); 
%%CITATION = NUPHZ,144,245;%%
%\cite{Czyz:2004rj}
%\bibitem{Czyz:2004rj}
H.~Czyz, A.~Grzelinska, J.~H.~Kuhn and G.~Rodrigo, 
%``The radiative return at Phi- and B-factories: FSR for muon pair  production
%at next-to-leading order,''
Eur.\ Phys.\ J.\ C \textbf{39}, 411 (2005) [arXiv:hep-ph/0404078]. 
%%CITATION = EPHJA,C39,411;%%
%\cite{Czyz:2004ua}
%\bibitem{Czyz:2004ua}
H.~Czyz, J.~H.~Kuhn, E.~Nowak and G.~Rodrigo, 
%``Nucleon form factors, B-meson factories and the radiative return,''
Eur.\ Phys.\ J.\ C \textbf{35}, 527 (2004) [arXiv:hep-ph/0403062]. 
%%CITATION = EPHJA,C35,527;%%
%\cite{Czyz:2002np}
%\bibitem{Czyz:2002np}
H.~Czyz, A.~Grzelinska, J.~H.~Kuhn and G.~Rodrigo, 
%``The radiative return at Phi- and B-factories: Small-angle photon emission
%at next to leading order,''
Eur.\ Phys.\ J.\ C \textbf{27}, 563 (2003) [arXiv:hep-ph/0212225]; 
%%CITATION = EPHJA,C27,563;
%\cite{Shwartz:2005tp}
%\bibitem{Shwartz:2005tp}

\bibitem{BBY} %\cite{Aubert:2005rm}
%\bibitem{Aubert:2005rm}
B.~Aubert \textit{et al.} [BABAR Collaboration], 
%``Observation of a broad structure in the $\pi^+ \pi^- J/\psi$ mass spectrum
%around 4.26-GeV/c$^2$,''
Phys.\ Rev.\ Lett.\ \textbf{95}, 142001 (2005) [arXiv:hep-ex/0506081]. 
%%CITATION = PRLTA,95,142001;%%

\bibitem{BBX} %\cite{Aubert:2006bu}
%\bibitem{Aubert:2006bu}
B.~Aubert \textit{et al.} [BABAR Collaboration], 
%``A structure at 2175-MeV in e+ e- --> Phi f0(980) observed via initial-state
%radiation,''
Phys.\ Rev.\ D \textbf{74}, 091103 (2006) [arXiv:hep-ex/0610018]; 
%%CITATION = PHRVA,D74,091103;%%X(2170)Babar
%\cite{Aubert:2007ur}
%\bibitem{Aubert:2007ur}
B.~Aubert \textit{et al.} [BABAR Collaboration], 
%``The $e^+ e^-\to K^+ K^- \pi^+\pi^-$, $K^+ K^- \pi^0\pi^0$ and $K^+ K^-
%K^+ K^-$ Cross Sections Measured with Initial-State Radiation,''
arXiv:0704.0630 [hep-ex]. %%CITATION = ARXIV:0704.0630;%%

\bibitem{KLOE} %\cite{Aulchenko:1998xy}
%\bibitem{Aulchenko:1998xy}
V.~M.~Aulchenko \textit{et al.} [SND Collaboration], 
%``First observation of Phi(1020) --> pi0 pi0 gamma decay,''
Phys.\ Lett.\ B \textbf{440}, 442 (1998) [arXiv:hep-ex/9807016]; 
%%CITATION = PHLTA,B440,442;%%
%\cite{Akhmetshin:1999di}
%\bibitem{Akhmetshin:1999di}
R.~R.~Akhmetshin et al. [CMD-2 Collaboration], 
%``Study of the Phi decays into pi0 pi0 gamma and eta pi0 gamma final  states,''
Phys.\ Lett.\ B \textbf{462}, 380 (1999) [arXiv:hep-ex/9907006]; 
%%CITATION = HEP-EX 9907006;%%
%\cite{Aloisio:2002bt}
%\bibitem{Aloisio:2002bt}
A.~Aloisio \textit{et al.} [KLOE Collaboration], 
%``Study of the decay Phi --> pi0 pi0 gamma with the KLOE detector,''
Phys.\ Lett.\ B \textbf{537}, 21 (2002) [arXiv:hep-ex/0204013]; 
%%CITATION = PHLTA,B537,21;%%
%\cite{Ambrosino:2006hb}
%\bibitem{Ambrosino:2006hb}
F.~Ambrosino \textit{et al.} [KLOE Collaboration], 
%``Dalitz plot analysis of e+ e- --> pi0 pi0 gamma events at s**(1/2) ~=
%M(Phi) with the KLOE detector,''
Eur.\ Phys.\ J.\ C \textbf{49}, 473 (2007) [arXiv:hep-ex/0609009]. 
%%CITATION = EPHJA,C49,473;%%
%\cite{Ambrosino:2005wk}
%\bibitem{Ambrosino:2005wk}
F.~Ambrosino \textit{et al.} [KLOE Collaboration], 
%``Study of the decay Phi --> f0(980) gamma --> pi+ pi- gamma with the  KLOE
%detector,''
Phys.\ Lett.\ B \textbf{634}, 148 (2006) [arXiv:hep-ex/0511031]. 
%%CITATION = PHLTA,B634,148;%%

\bibitem{MSL} %\cite{Bramon:2002iw}
%\bibitem{Bramon:2002iw}
A.~Bramon, R.~Escribano, J.~L.~Lucio M, M.~Napsuciale and G.~Pancheri, 
%``Scalar f0(980) and sigma(500) meson exchange in Phi decays into  pi0 pi0
%gamma,''
Eur.\ Phys.\ J.\ C \textbf{26}, 253 (2002) [arXiv:hep-ph/0204339]; 
%%CITATION = EPHJA,C26,253;%%
%\cite{Black:2006mn}
%\bibitem{Black:2006mn}
D.~Black, M.~Harada and J.~Schechter, 
%``Chiral approach to Phi radiative decays,''
Phys.\ Rev.\ D \textbf{73}, 054017 (2006) [arXiv:hep-ph/0601052]; 
%%CITATION = PHRVA,D73,054017;%%
%\cite{Black:2002ek}
%\bibitem{Black:2002ek}
D.~Black, M.~Harada and J.~Schechter, 
%``Vector meson dominance model for radiative decays involving light  scalar
%mesons,''
Phys.\ Rev.\ Lett.\ \textbf{88}, 181603 (2002) [arXiv:hep-ph/0202069]. 
%%CITATION = PRLTA,88,181603;%%

\bibitem{MHOT} %%CITATION = PHLTA,B426,7;%%
%\cite{Marco:1999df}
%\bibitem{Marco:1999df}
E.~Marco, S.~Hirenzaki, E.~Oset and H.~Toki, 
%``Radiative decay of rho0 and Phi mesons in a chiral unitary approach,''
Phys.\ Lett.\ B \textbf{470}, 20 (1999) [arXiv:hep-ph/9903217]; 
%%CITATION = PHLTA,B470,20;%%
%\cite{Markushin:2000fa}
%\bibitem{Markushin:2000fa}
V.~E.~Markushin, 
%``The radiative decay Phi --> gamma pi pi in a coupled channel model and  the
%structure of f0(980),''
Eur.\ Phys.\ J.\ A \textbf{8}, 389 (2000) [arXiv:hep-ph/0005164]; 
%%CITATION = EPHJA,A8,389;%%
%\cite{Palomar:2003rb}
%\bibitem{Palomar:2003rb}
J.~E.~Palomar, L.~Roca, E.~Oset and M.~J.~Vicente Vacas, 
%``Sequential vector and axial-vector meson exchange and chiral loops in
%radiative Phi decay,''
Nucl.\ Phys.\ A \textbf{729}, 743 (2003) [arXiv:hep-ph/0306249]; 
%%CITATION = NUPHA,A729,743;%%
%\cite{Oller:2002na}
%\bibitem{Oller:2002na}
  J.~A.~Oller,
  %``Finite width effects in Phi radiative decays,''
  Nucl.\ Phys.\  A {\bf 714}, 161 (2003)
  [arXiv:hep-ph/0205121].
  %%CITATION = NUPHA,A714,161;%%
  
\bibitem{Oller} %\cite{Oller:1998ia}
%\bibitem{Oller:1998ia}
J.~A.~Oller, %``The Phi --> gamma K0 anti-K0 decay,''
Phys.\ Lett.\ B \textbf{426}, 7 (1998) [arXiv:hep-ph/9803214].

\bibitem{OOP} %\cite{Oller:2000ug}
%\bibitem{Oller:2000ug}
J.~A.~Oller, E.~Oset and J.~E.~Palomar, 
%``Pion and kaon vector form factors,''
Phys.\ Rev.\ D \textbf{63}, 114009 (2001) [arXiv:hep-ph/0011096]. 
%%CITATION = PHRVA,D63,114009;%%

\bibitem{BiselloFF} %\cite{Bisello:1988ez}
%\bibitem{Bisello:1988ez}
D.~Bisello \textit{et al.} [DM2 Collaboration], 
%``Study Of The Reaction E+ E- $\to$ K+ K- In The Energy Range 1350 <=
%S**(1/2) <= 2400-Mev,''
Z.\ Phys.\ C \textbf{39}, 13 (1988); %%CITATION = ZEPYA,C39,13;%%
%\cite{Delcourt:1980eq}
%\bibitem{Delcourt:1980eq}
B.~Delcourt, D.~Bisello, J.~C.~Bizot, J.~Buon, A.~Cordier and F.~Mane, 
%``Study Of The Reaction E+ E- $\to$ K+ K- In The Total Energy Range 1400-Mev
%To 2060-Mev,''
Phys.\ Lett.\ B \textbf{99}, 257 (1981). %%CITATION = PHLTA,B99,257;%%
%\cite{Mane:1980ep}
%\bibitem{Mane:1980ep}
F.~Mane, D.~Bisello, J.~C.~Bizot, J.~Buon, A.~Cordier and B.~Delcourt, 
%``Study Of The Reaction E+ E- $\to$ K0(S) K0(L) In The Total Energy Range
%1.4-Gev To 2.18-Gev And Interpretation Of The K+ And K0 Form-Factors,''
Phys.\ Lett.\ B \textbf{99}, 261 (1981); %%CITATION = PHLTA,B99,261;%%

\bibitem{BiselloCS} %\cite{Mane:1982si}
%\bibitem{Mane:1982si}
F.~Mane, D.~Bisello, J.~C.~Bizot, J.~Buon, A.~Cordier and B.~Delcourt, 
%``Study Of E+ E- $\to$ K0(S) K+- Pi-+ In The 1.4-Gev To 2.18-Gev Energy
%Range: A New Observation Of An Isoscalar Vector Meson Phi-Prime (1.65-Gev),''
Phys.\ Lett.\ B \textbf{112}, 178 (1982). %%CITATION = PHLTA,B112,178;%%

\bibitem{EGPR} %\cite{Ecker:1988te}
%\bibitem{Ecker:1988te}
G.~Ecker, J.~Gasser, A.~Pich and E.~de Rafael, 
%``The Role Of Resonances In Chiral Perturbation Theory,''
Nucl.\ Phys.\ B \textbf{321}, 311 (1989). %%CITATION = NUPHA,B321,311;%%

\bibitem{OO} %\cite{Oller:1997ti}
%\bibitem{Oller:1997ti}
J.~A.~Oller and E.~Oset, 
%``Chiral symmetry amplitudes in the S-wave isoscalar and isovector  channels
%and the sigma, f0(980), a0(980) scalar mesons,''
Nucl.\ Phys.\ A \textbf{620}, 438 (1997) [Erratum-ibid.\ A \textbf{652}, 407
(1999)] [arXiv:hep-ph/9702314].

\bibitem{ND} %\cite{Oller:1998zr}
%\bibitem{Oller:1998zr}
J.~A.~Oller and E.~Oset, 
%``N/D description of two meson amplitudes and chiral symmetry,''
Phys.\ Rev.\ D \textbf{60}, 074023 (1999) [arXiv:hep-ph/9809337]. 
%%CITATION = PHRVA,D60,074023;%%

\bibitem{Ramos} %\cite{Oset:1997it}
%\bibitem{Oset:1997it}
E.~Oset and A.~Ramos, 
%``Non perturbative chiral approach to s-wave anti-K N interactions,''
Nucl.\ Phys.\ A \textbf{635}, 99 (1998) [arXiv:nucl-th/9711022]. 
%%CITATION = NUPHA,A635,99;%%

\bibitem{Ulf} %\cite{Oller:2000fj}
%\bibitem{Oller:2000fj}
J.~A.~Oller and U.~G.~Meissner, 
%``Chiral dynamics in the presence of bound states: Kaon nucleon  interactions
%revisited,''
Phys.\ Lett.\ B \textbf{500}, 263 (2001) [arXiv:hep-ph/0011146]. 
%%CITATION = PHLTA,B500,263;%%

\bibitem{OOPel} %%CITATION = NUPHA,A620,438;%%
%\cite{Oller:1997ng}
%\bibitem{Oller:1997ng}
J.~A.~Oller, E.~Oset and J.~R.~Pelaez, 
%``Non-perturbative approach to effective chiral Lagrangians and meson
%interactions,''
Phys.\ Rev.\ Lett.\ \textbf{80}, 3452 (1998) [arXiv:hep-ph/9803242]; 
%%CITATION = PRLTA,80,3452;%%
%\cite{Oller:1998hw}
%\bibitem{Oller:1998hw}
J.~A.~Oller, E.~Oset and J.~R.~Pelaez, 
%``Meson meson and meson baryon interactions in a chiral non-perturbative
%approach,''
Phys.\ Rev.\ D \textbf{59}, 074001 (1999) [Erratum-ibid.\ D \textbf{60},
099906 (1999)] [arXiv:hep-ph/9804209]. %%CITATION = PHRVA,D59,074001;%%

\bibitem{Pich} %\cite{Rosell:2004mn}
%\bibitem{Rosell:2004mn}
I.~Rosell, J.~J.~Sanz-Cillero and A.~Pich, 
%``Quantum loops in the resonance chiral theory: The vector form factor,''
JHEP \textbf{0408}, 042 (2004) [arXiv:hep-ph/0407240]. 
%%CITATION = JHEPA,0408,042;%%

\bibitem{GVFV}
%\cite{Ecker:1989yg}
%\bibitem{Ecker:1989yg}
  G.~Ecker, J.~Gasser, H.~Leutwyler, A.~Pich and E.~de Rafael,
  %``Chiral Lagrangians for Massive Spin 1 Fields,''
  Phys.\ Lett.\  B {\bf 223}, 425 (1989).
  %%CITATION = PHLTA,B223,425;%%
\bibitem{structure}  
%\cite{Ding:2006ya}
%\bibitem{Ding:2006ya}
  G.~J.~Ding and M.~L.~Yan,
  %``A candidate for 1- strangeonium hybrid,''
  Phys.\ Lett.\  B {\bf 650}, 390 (2007)
  [arXiv:hep-ph/0611319];
  %%CITATION = PHLTA,B650,390;%%
%\cite{Ding:2007pc}
%\bibitem{Ding:2007pc}
  G.~J.~Ding and M.~L.~Yan,
  %``Y(2175): Distinguish hybrid state from higher quarkonium,''
  arXiv:hep-ph/0701047;
  %%CITATION = HEP-PH/0701047;%%  
  %\cite{Wang:2006ri}
%\bibitem{Wang:2006ri}
  Z.~G.~Wang,
  %``Analysis of Y(2175) as a tetraquark state with QCD sum rules,''
  Nucl.\ Phys.\  A {\bf 791}, 106 (2007)
  [arXiv:hep-ph/0610171].
  %%CITATION = NUPHA,A791,106;%%
\end{thebibliography}
\end{document}